%% file: HS-RS.tex
\documentclass[letterpaper,11pt]{article}
\usepackage[DIV12]{typearea}
\usepackage{longtable}
\usepackage{booktabs,colortbl}
\usepackage{multirow}
\usepackage{amsmath}
\usepackage{amssymb}
\usepackage{bbm}
\usepackage[utf8]{inputenc}
\usepackage{pdflscape}
\usepackage{xspace}
\usepackage[caption=false]{subfig}
\usepackage{nicefrac}
\usepackage{graphicx}
\usepackage{cite}  
\usepackage[small]{caption2}
\usepackage{mathtools}
\usepackage{nccfoots}
\usepackage{dsfont}
\usepackage{braket}
\usepackage{tikz}
\usetikzlibrary{calc,positioning}
\usepackage[textsize=tiny,backgroundcolor=yellow]{todonotes}

\usepackage[all,cmtip]{xy}
\newlength{\xywd}
\newcommand{\xyrightarrow}[2][]{%
  \sbox{0}{$\scriptstyle#1$}%
  \xywd=\wd0
  \sbox{0}{$\scriptstyle#2$}%
  \ifdim\wd0>\xywd \xywd=\wd0 \fi
  \xymatrix@C\dimexpr\xywd+1em\relax{{}\ar[r]^{#2}_{#1}&{}}%
}

\renewcommand{\thefootnote}{\alph{footnote}}

\newcommand{\rep}[1]{\ensuremath\boldsymbol{#1}}
\newcommand{\crep}[1]{\ensuremath\overline{\boldsymbol{#1}}}
\newcommand{\Z}[1]{\ensuremath{\mathds{Z}_{#1}}} 
\newcommand{\SO}[1]{\ensuremath{\mathrm{SO}(#1)}}
\newcommand{\SU}[1]{\ensuremath{\mathrm{SU}(#1)}}
\newcommand{\SL}[1]{\ensuremath{\mathrm{SL}(#1)}}
\newcommand{\GL}[1]{\ensuremath{\mathrm{GL}(#1)}}
\newcommand{\U}[1]{\ensuremath{\mathrm{U}(#1)}}
\newcommand{\E}[1]{\ensuremath{\mathrm{E}_{#1}}}
\newcommand{\e}{\mathrm{e}}
\newcommand{\I}{\mathrm{i}}
\newcommand{\Id}{\mathds{1}}

\newcommand{\x}{\ensuremath{\times}}

\usepackage{xcolor}
\definecolor{darkgreen}{HTML}{109930}
\definecolor{pink}{rgb}{0.858, 0.188, 0.478}

\advance \headheight by 3.0truept       
\usepackage[pdftex,hidelinks,colorlinks=true,citecolor=darkgreen]{hyperref}
\hypersetup{
    pdftitle = {Non-Abelian orbifolds of the SO(32) heterotic string},
    pdfauthor = {Hernandez-Segura, Ramos-Sanchez}
}

\usepackage[toc,page]{appendix}    

\usepackage{cleveref}

\begin{document}

\begin{titlepage}

\begin{flushright}
\end{flushright}

\vspace*{1.0cm}

\begin{center}
{\Large\textbf{\boldmath Non-Abelian orbifolds of the \SO{32} heterotic string\unboldmath}}

\vspace{1cm}
\textbf{Miguel Hern\'andez--Segura\footnote{\texttt{hernandez@estudiantes.fisica.unam.mx}}}
and
\textbf{Sa\'ul Ramos--S\'anchez\footnote{\texttt{ramos@fisica.unam.mx}}}\\[5mm]

\textit{\small Instituto de F\'isica, Universidad Nacional Aut\'onoma de M\'exico,\\ POB 20-364, Cd.Mx. 01000, M\'exico}
\end{center}

\vspace{1cm}

\vspace*{1.0cm}

\begin{abstract}
Non-Abelian toroidal heterotic orbifolds have received comparatively little attention, mainly because of the significant
computational challenges they pose, even at the level of computing their matter spectrum.
Similarly, the \SO{32} heterotic string remains relatively unexplored. In this paper, we provide some useful
tools to handle this situation. We find that certain non-Abelian orbifolds can be studied using the techniques
that are common to Abelian compactifications. In such cases, we show how to compute the gauge groups and massless
matter spectrum for non-Abelian orbifolds of the \SO{32} heterotic string with standard embedding.
A general feature of these constructions is the reduction of the rank of the gauge group.
Our findings motivate further research on non-Abelian orbifolds with non-standard embedding,
where realistic, rank-reduced models are expected to emerge.

\end{abstract}

\end{titlepage}

\newpage
\setcounter{footnote}{0} 
\renewcommand{\thefootnote}{\arabic{footnote}}

\section{Introduction}

String model building has led to many appealing results in particular based on compactifications of the heterotic string
on Abelian toroidal orbifolds~\cite{Bailin:1999nk,Lebedev:2006kn,Olguin-Trejo:2018wpw,Baur:2019iai,Baur:2022hma,Ramos-Sanchez:2024keh},
which can even be generated automatically~\cite{Nilles:2011aj,Escalante-Notario:2025hvn}.
However, non-Abelian toroidal orbifolds have been historically disregarded mainly because of computational complications
associated with identifying the properties of the effective limit after compactification.

One important reason for studying non-Abelian orbifolds was given in the bottom-up approach~\cite{Hebecker:2003}:
non-Abelian twists reduce the rank of the gauge group of the original higher dimensional theory. Such twists
are naturally embedded in non-Abelian toroidal orbifolds of string theory. In the top-down approach, there have
been some important efforts to shed light on toroidal non-Abelian orbifolds. In particular, after the classification
of all admissible six-dimensional (6D) toroidal orbifolds of ref.~\cite{Fischer:2012qj}, the first steps to study
the \E8\x\E8 heterotic string compactified on non-Abelian orbifolds have been given. Using the Hodge numbers of
non-Abelian orbifolds have helped to determine the resulting $\rep{27}$ and $\crep{27}$ gauge representations of
\E6 in orbifold compactifications with standard embedding~\cite{Fischer:2013qza}. Also, some techniques have been
developed in ref.~\cite{Konopka:2012gy} to arrive at the four-dimensional (4D) matter spectrum of the \E8\x\E8
heterotic string compactified on an $S_3$ non-Abelian orbifold.

On the other hand, the \SO{32} heterotic string has been much less explored perhaps due to the lack of spinorial
representations in ten dimensions, which makes it harder to lead to spinors in four dimensions. Nevertheless,
many models based on the \SO{32} heterotic string compactified on Abelian toroidal orbifolds have been
built~\cite{Giedt:2003an,Giedt:2004wd,Choi:2004wn,Blumenhagen:2005pm,Nilles:2006np,Kim:2020erx} 
and even semi-realistic models with three standard-model generations have been found~\cite{Ramos-Sanchez:2008nwx} 
(see also~\cite{Abe:2015mua,Abe:2015xua} for similar approaches). 

This work contributes to the development of string phenomenology based on non-Abelian orbifold compactifications
of the \SO{32} heterotic string. We address the extension of traditional techniques used to compute the matter
spectrum in Abelian orbifolds to the non-Abelian case. We focus on standard embedding of 6D orbifolds into
the gauge degrees of freedom.

\enlargethispage{\baselineskip}
Our work is organized as follows. First, in section~\ref{sec:SO32} we provide an overview of the \SO{32}
heterotic string. Next, in section~\ref{sec:orbifolds} we define orbifolds and present the main aspects to consider when
working with them in the Abelian scenario; also a brief description of certain generic features of non-Abelian orbifolds is given.
Then, in section~\ref{sec:rankr}, we elaborate on the idea of rank reduction in non-Abelian orbifold compactifications.
In section~\ref{sec:spectrum}, we discuss first how topological information can help us find all non-trivial
representations under the \SO{26} gauge factor in 4D of all non-Abelian heterotic orbifolds. Then, we explain in detail how to generalize the
standard Abelian techniques to determine the complete massless matter spectrum of certain non-Abelian
orbifolds. We use $S_3$ as an explicit example. After this, in section~\ref{sec:results}
we present our results for the orbifolds with $D_{4}$ and $(\Z{4}\x\Z{2})\rtimes\Z{2}$ space groups. Finally, in
section~\ref{sec:conclusions}, we present our conclusions and discuss some future directions for this work. Our appendices
provide useful algebraic elements and a list of our results for all non-Abelian orbifolds.

\input{so32string}

\input{orbifolds}

\input{rankreduction}

\input{spectrum}

\input{results}

\section{Conclusions and Outlook}
\label{sec:conclusions}

In this work we provided tools to 4D massless matter spectrum of non-Abelian orbifold compactifications of the \SO{32} heterotic string
with standard embedding. First, using results obtained in Abelian orbifolds relying purely on their topological aspects, we have found the
nontrivial matter representations of \SO{26} of all admissible non-Abelian orbifolds. These results are presented in Appendix~\ref{app:26s}.

More interestingly, we have computed the complete 4D massless matter spectrum of a class of non-Abelian \SO{32} heterotic orbifolds
with standard embedding, based on the formalism for Abelian orbifolds. This extension of the Abelian techniques consists in an
algorithm that allows us to block diagonalize orbifold twists in all the various sectors of the compactification,
along with a method to arrive at the effective invariant states under the orbifold. The identification of the matter states allows us
to compute all and their gauge transformations, including \U1 charges. We verify the consistency of this method by
i) confirming that the complete spectrum is anomaly free, and ii) comparing the number of $\rep{26}$ of \SO{26}
obtained with this procedure and the topological technique.

As expected from bottom-up discussions, we find that rank reduction is a general feature of the effective theory of non-Abelian
heterotic orbifolds with standard embedding.
In this case, the gauge group always includes an \SO{26} unbroken gauge group, which can be accompanied by a gauge factor such as $\U1^2$ or \U1 or nothing.
Even though standard
embedding only allows us to reduce the rank by up to three units, it is conceivable that non-standard embedding can provide models
with further rank reduction and, hence, with more realistic features. This shall be done elsewhere.

There are several additional questions left to answer in future works. For example, some orbifold geometries seem to be incompatible with
our methods. Hence, we must try and find an alternative mechanism to compute the massless spectrum in such cases.
Independently of this puzzle, one might wonder what the geometric and modular symmetries relating twisted states
can be in these non-Abelian geometries. Such {\it flavor} symmetries might be found by using known techniques
based on the Narain formalism, as in the Abelian case~\cite{Nilles:2020kgo,Nilles:2020tdp,Nilles:2020gvu,Baur:2020jwc,Baur:2021mtl}.

\subsection*{Acknowledgments}
This work was partly supported by UNAM-PAPIIT grant IN113223 and Marcos Moshinsky Foundation.
It is a pleasure to thank Sebastian Konopka for useful discussions.

\appendix

\input{appendix}


\newpage

{\small

\providecommand{\bysame}{\leavevmode\hbox to3em{\hrulefill}\thinspace}

}
\end{document}

%% file: so32string.tex
\section{\boldmath \SO{32} heterotic string}
\label{sec:SO32}

The heterotic string theories describe $\mathcal N=1$ supergravity in a 10D flat spacetime. The fermionic
and bosonic states are constructed from right and left-moving closed-string modes, related by supersymmetric
transformations. Additionally, these theories feature 16 extra dimensions corresponding to gauge degrees of freedom.
Consistency requirements dictate that these dimensions must be compactified on a torus $\mathds{T}^{16}=\mathds{R}^{16}/\Lambda$,
where $\Lambda$ is one of the only two admissible 16-dimensional (16D), even and self-dual lattices: the
root lattice of \E8\x\E8 or the weight lattice of $\mathrm{Spin}(32)/\Z2$. Depending on the choice of $\Lambda$,
the resulting 10D theory exhibits either the \E8\x\E8  or the \SO{32} gauge group.

We focus on the heterotic string with \SO{32} gauge group.
In the bosonic formulation, the properties of the massless states of the \SO{32} heterotic string arise from the
solutions of the masslessness (and level-matching) conditions,
\begin{equation}
\label{eq:masslessSO32}
 \frac12 p^2 + \tilde N - 1 ~=~ 0 ~=~ \frac12 q^2 -\frac12\,,
\end{equation}
where $p$ denotes the left-moving gauge momentum of the string, which takes values in the 16D self-dual weight lattice of
$\mathrm{Spin}(32)/\Z2$. Similarly, $q$ denotes the right-moving momenta of the string, taking values in
spatial weight lattice of the little Lorentz group \SO{8} in 10D.\footnote{A basis of the \SO{8} and $\mathrm{Spin}(32)/\Z2$
weight lattices is presented in Appendix~\ref{app:Lattices}.} $\tilde N$ is a number operator that counts the
left-moving oscillators $\tilde\alpha_{-n}$, $n\in\mathds{N}$, acting on the left-moving bosonic vacuum $\ket{0}_L$,
i.e.\footnote{Here, $\tilde\alpha^\mu_{n}\ket0_L=0$ and $\tilde\alpha^\mu_{-n}$ are creation operators for $n\in\mathds{N}$.}
\begin{equation}
\label{eq:oscNumber}
 \tilde N := \sum_{n=1}^\infty \tilde\alpha_{-n}\cdot\tilde\alpha_{n}\,,\quad \text{with}\quad
 \tilde\alpha_{-n}\cdot\tilde\alpha_{n} := \sum_{i=1}^8\tilde\alpha_{-n}^i\tilde\alpha_{n}^i +
                                             \sum_{I=1}^{16}\tilde\alpha_{-n}^I\tilde\alpha_{n}^I\,,\;
[\tilde\alpha_n^{\mu},\tilde\alpha_m^{\nu}] = n \delta^{\mu\nu}\delta_{n+m,0} \,.
\end{equation}
In the scalar product, we consider all dynamical degrees of freedom $x^\mu$ of left-moving oscillators, which include
the 16 gauge degrees of freedom $x^I$, and the eight light-cone spatial degrees of freedom $x^i$,
which are adequate for massless states.
Hence, the massless spectrum of the \SO{32} heterotic string comprises
\begin{subequations}
\begin{align}
  \tilde\alpha^i_{-1}\ket{0}_L  \otimes \ket{q}_R  &:&&  \mathcal N=1 \text{ supergravity multiplet}\,,\\
  \tilde\alpha^I_{-1}\ket{0}_L  \otimes \ket{q}_R  &:&&   \text{16 \SO{32} neutral gauge bosons and gauginos}\,,\\
  \ket{p} \otimes \ket{q}_R                   &:&&  \text{480 \SO{32} charged gauge bosons and gauginos}\,,
\end{align}
\end{subequations}
with $p^2=2$ and $q^2=1$ (see Appendix~\ref{app:Lattices} for the explicit form of $p$ and $q$).
One can show that the latter two sets of
states build the full super-Yang-Mills multiplet with \SO{32} gauge group, where gauge bosons transform
in the $\rep{496}$ representation of \SO{32}.

Group-theoretically, the 480 charged gauge bosons in $\ket{p}_{L} \otimes \ket{q}_R$ can be interpreted as
ladder operators of the orthogonal group \SO{32} of rotations in 32 dimensions. In these terms, the
16 neutral gauge bosons correspond to the generators of the \SO{32} Cartan subalgebra, which can be
expressed as
\begin{equation}
\label{eq:CartanSO32}
  H_{\SO{32}} ~=~ \{J_{1,2},J_{3,4},\ldots, J_{31,32}\}\,.
\end{equation}
Here, $J_{M,N}$ denotes the generator of rotations in the $M-N$ coordinate plane of the gauge degrees of freedom.
It is important to stress that~\eqref{eq:CartanSO32} represents a specific basis choice and, as we shall see,
in some cases, a different choice may be required.

%% file: orbifolds.tex
\section{Heterotic orbifolds}
\label{sec:orbifolds}

Assuming the existence of $\mathcal N=1$ supersymmetry at low energies, we must demand that the \SO{32} heterotic string
be compactified on either a Calabi-Yau manifold or one of the 138 admissible toroidal orbifold geometries~\cite{Fischer:2012qj}.
Toroidal orbifolds are simple spaces in the sense that they are
flat everywhere except for a finite set of curvature singularities or fixed points, so that e.g.\ a (flat) metric of the
compact space can be properly defined (almost) everywhere in the orbifold, and standard CFT (and QFT) techniques can be used
to determine the properties of the effective 4D field theory after compactification.

A general toroidal orbifold of a heterotic string can be split as
\begin{equation}
  \mathcal{O} ~=~ \mathcal{O}_6 \otimes \mathcal{O}_{\mathrm{gauge}}\,,
\end{equation}
where $\mathcal{O}_6$ and $\mathcal{O}_{\mathrm{gauge}}$ denote toroidal orbifolds of the six extra dimensions and
the gauge degrees of freedom, respectively, i.e.
\begin{equation}
  \mathcal{O}_6 ~:=~ \mathds{T}^6 / P
  \qquad\text{and}\qquad
  \mathcal{O}_{\mathrm{gauge}} ~:=~ \mathds{T}^{16}/\mathcal G\,.
\end{equation}
$P$ is known as the 6D point group and corresponds to a set of discrete isometries of a torus $\mathds{T}^6$, which can be chosen
from the classification of ref.~\cite{Fischer:2012qj}. If $P$ is a non-Abelian (Abelian) group, the orbifold is said to be non-Abelian (Abelian).
$\mathcal G$ is the so-called gauge-twisting group and is a gauge embedding of $P$ in the 16 gauge degrees of freedom,
$P \hookrightarrow \mathcal G$, subject to modular-invariance conditions~\cite{Dixon:1986jc,Vafa:1986wx,Ploger:2007iq}.
$\mathcal G$ acts on $\mathds{T}^{16}=\mathds{R}^{16}/\Lambda$, where $\Lambda$ is the weight lattice of $\mathrm{Spin}(32)/\Z2$.

\subsection{6D orbifolds: the space group}
\label{sec:GeometricOrbi}

The 6D torus $\mathds{T}^6$ of the orbifold $\mathcal{O}_6$ is spanned by a lattice $\Gamma$, such that
\begin{equation}
\mathds T^6 ~=~ \mathds{R}^{6}/\Gamma \qquad\text{with}\qquad
\Gamma ~=~ \left\{ n_{\alpha}e_{\alpha}~|~n_{\alpha}\in\Z{}, \alpha=1,2,\ldots,6 \right\}\,.
\end{equation}
where $\Gamma$ is left invariant by $P$ because it is a discrete isometry group of the torus.
Considering this, the orbifold can equivalently be defined as the quotient
\begin{equation}
\label{eq:defOrbifoldinS}
 \mathcal{O}_6 ~=~ \mathds{R}^6/S
\end{equation}
in terms of the so-called space group $S$, whose elements $g:=(\vartheta,\lambda)\in S$ include the action of the point group
$\vartheta\in P$ and a translation $\lambda = q_{\alpha}e_{\alpha}$, where $e_\alpha\in\Gamma$ and $q_\alpha$ are rationals.
Note that, only if $q_\alpha=n_\alpha\in\Z{}$, then $\lambda\in\Gamma$. If $q_\alpha\not\in\Z{}$, $g\in S$ is called a roto-translation.
The space group is given by $S = P\ltimes\Gamma$ when it includes no roto-translations.

The action of $g=\left(\vartheta,\lambda \right)$ on $X\in\mathds{R}^6$ is given by
\begin{equation}
\label{eq:actionSonX}
  X ~\xmapsto{~g~}~ g X ~:=~ \vartheta\, X + \lambda\,.
\end{equation}
Hence, the 6D orbifold~\eqref{eq:defOrbifoldinS} is explicitly given by the identification
\begin{equation}
  X ~\sim~ \vartheta\, X + \lambda\,,
  \qquad \forall\quad\vartheta\in P\,,\quad \lambda = q_{\alpha}e_{\alpha}\,.
\end{equation}
Further, eq.~\eqref{eq:actionSonX} implies that space-group elements satisfy
\begin{subequations}
\begin{align}
g\,h    &=\left(\vartheta_{1}\vartheta_{2}, \lambda_1 + \vartheta_{1}\lambda_2\right),& g,h\,\in S, \\
g^{-1}  &=\left(\vartheta,\lambda\right)^{-1}=\left(\vartheta^{-1},-\vartheta^{-1}\lambda\right),& g\,\in S
\end{align}
\end{subequations}
with $g^{-1}g=g g^{-1} = (\Id,0)$.

Even though clearly all space groups $S$ are non-Abelian, it is customary to consider orbifolds to be Abelian or non-Abelian,
depending entirely on the nature of the corresponding point group $P$, as mentioned before. In the following we discuss separately both cases.

\subsection{Abelian orbifolds}
\label{sec:AOrbi}

The Abelian point groups that give rise to Abelian orbifolds have a simple form: they are restricted to be either cyclic finite groups or direct
products of two cyclic groups,
\begin{equation}
P ~\in~ \left\{ \Z{N},\,\,\Z{N}\x\Z{M} \right\}.
\end{equation}
All admissible Abelian point groups leading to $\mathcal N=1$ SUSY in 4D are presented in table~\ref{tab:aoN1}.

\begin{table}[b]
\begin{center}
\begin{tabular}{|c|c|c|c|c|}
\hline
$\Z{3}$        & $\Z{4}$        & $\Z{6}$-I        & $\Z{6}$-II        & $\Z{7}$ \\
\hline
$\Z{8}$-I      & $\Z{8}$-II     & $\Z{12}$-I       & $\Z{12}$-II       & $\Z{2}\x\Z{2}$\\
\hline
$\Z{2}\x\Z{4}$ & $\Z{3}\x\Z{3}$ & $\Z{2}\x\Z{6}$-I & $\Z{2}\x\Z{6}$-II & $\Z{4}\x\Z{4}$ \\
\hline
$\Z{3}\x\Z{6}$ & $\Z{6}\x\Z{6}$ \\
\cline{1-2}
\end{tabular}
\caption{\label{tab:aoN1}
All 17 Abelian point groups compatible with 4D $\mathcal{N}=1$ SUSY, which yield
138 inequivalent orbifold geometries~\cite{Fischer:2012qj}.}
\end{center}
\end{table}

Given a choice of an orbifold point group, we must define the action of this group on the spacetime coordinates.
In general, it is convenient to rewrite the orbifold action on the real coordinates $X^i$, $i\in\{3,\ldots, 8\}$,
as acting on the three complexified dimensions given by
\begin{equation}
Z^{a} ~:=~ \frac{1}{\sqrt{2}}\left( X^{2a+1}+\I X^{2a+2} \right)\,,\qquad a\in\{ 1,2,3 \}\,.
\end{equation}
In this basis, $P$ elements can be represented as diagonal $3\x3$ matrices. Let us consider as our working example for
this section the case of $P=\Z{N}$. The point group generator, called twist, can be expressed as
\begin{equation} \label{eq:gent}
\theta ~=~\operatorname{diag}\left( \e^{2\pi \I v^{1}},  \e^{2\pi \I v^{2}},  \e^{2\pi \I v^{3}} \right),
\end{equation}
where $v=\left( v^{0}=0, v^{1}, v^{2}, v^{3} \right)$ is the so-called twist vector. In this notation, we have introduced the trivial twist
component $v^0=0$ in order for $\theta$ to include a trivial action on $Z^0:=\frac{1}{\sqrt2} (X^1 + \I X^2)$, which describes the two real
light-cone coordinates of our 4D spacetime.
The action on all coordinates is then given by $\mathrm{diag}(1,\theta)$, expressed in the (canonical) Cartan basis of \SO{8},
\begin{equation}
  H_{\SO8} ~=~ \left\{ J_{1,2},\,J_{3,4},\,J_{5,6},\,J_{7,8} \right\},
\end{equation}
where $J_{i,j}$ is the generator of rotations in the $X^{i}-X^{j}$ plane. This implies that all point-group
elements can be generated by the \SO6 Cartan subalgebra $H_{\SO6}=\{J_{3,4}, J_{5,6},J_{7,8}\}$.

In order to arrive at 4D $\mathcal{N}=1$ SUSY, preserving only one 4D gravitino invariant, it is known
that the holonomy group of the orbifold must satisfy that $\operatorname{Hol}\left(
\mathcal{O}_{6} \right)\subset\SU{3}$. This imposes the condition $P\subset \SU{3}$, which,
in terms of the twist vector $v$, becomes the condition
\begin{equation}
\pm v^{1}\pm v^{2}\pm v^{3} ~=~ 0\,.
\end{equation}
For the point groups included in table~\ref{tab:aoN1} it is possible to find a twist vector satisfying this constraint.

As mentioned earlier, an orbifold exhibits curvature singularities identified as fixed points of the orbifold.
Fixed points are locations $Z_f$ in the compact dimensions, where the action of the elements of the
space group is trivial. E.g.\ for $g=(\theta^k,\lambda) \in S$, then $Z_f$ is a 6D fixed point if it satisfies
\begin{equation}
  Z_f ~\xmapsto{~g~} g\, Z_f ~=~ \theta^k Z_f + \lambda ~\stackrel!=~ Z_f\,.
\end{equation}
This implies that all fixed points associated with $g$ can be expressed as
\begin{equation}
\label{eq:fixedpoints1}
  Z_f ~=~ (\Id - \theta^k)^{-1} \lambda\,,
\end{equation}
provided that $(\Id - \theta^k)$ is non-singular. If it is singular, one can always extract a non-singular
diagonal block to determine the fixed points in less than six dimensions (typically four), and the
singular part defines a fixed (typically 2D) subspace. We obtain this way so-called fixed tori.
Note that $\lambda$ is defined in the torus up to $\Gamma$ lattice translations, i.e.\ $\lambda$ and $\lambda + \gamma$,
$\gamma\in\Gamma$, are equivalent. Hence, in the absence of roto-translations, it suffices to consider
elements $\theta^k$ of the point group $P$. Inequivalent powers $k$ can lead to different fixed points or tori.
For this reason, it is customary to consider each power of $k$ a different {\it sector} of the orbifold,
where the associated fixed points are obtained from eq.~\eqref{eq:fixedpoints1} with $g=(\theta^k,0)$
and $\lambda=\gamma\in\Gamma$. Similarly, for orbifolds with roto-translations, one takes the various
powers of the $S$ generators $g=(\theta,\lambda)$, $\lambda\not\in\Gamma$, as different sectors, and all
associated fixed points are obtained by considering $\lambda=\lambda+\gamma$ with various $\gamma\in\Gamma$.
Up to elements of the null space of $(\Id - \theta^k)$, we can define $g$ with a fixed $\lambda$
as the {\it constructing element} associated with a fixed point $Z_f$.
As we shall see in section~\ref{sec:spectrum}, the centralizer of the constructing elements are useful.
The centralizer of $g$ in $S$ is defined by
\begin{equation}
\label{eq:centralizer}
  \mathcal{C}_{S}(g) ~:=~ \left\{ h\in S ~|~ [g,h]=0 \right\}\,.
\end{equation}

Next, we embed the geometric action of the orbifold into the 16D \SO{32} gauge degrees of freedom.
In the standard formalism~\cite{Bailin:1999nk,Ramos:2008pt,Vaudrevange:2008sm}, the most straightforward
approach is a shift embedding, wherein the elements of $S$ are mapped to 16D shift vectors, effectively
``translating'' the orbifold action into the gauge sector. Explicitly,
\begin{equation}
 \label{eq:encaje1}
 \left( \theta^{k},n_{\alpha}e_{\alpha} \right) ~\hookrightarrow~ \left( kV,n_{\alpha}A_{\alpha} \right)\,,\qquad
 0 \leq k < N\,,
\end{equation}
where $k,\, n_{\alpha}\,\in\Z{}$, $V$ is the shift vector, $A_{\alpha}$ denote so-called Wilson lines,
and $\theta^k$ is a generic point-group element for $P=\Z{N}$. $( kV,n_{\alpha}A_{\alpha})$ is then an element
of the gauge twisting group $\mathcal G$, whose action on the gauge left-moving momenta $p$ consists in shifting them
as $p \mapsto p_{sh} := p + kV + n_{\alpha}A_{\alpha}$. Frequently, the shift is related to the original space-group
element $s$ by defining $V_g := kV + n_{\alpha}A_{\alpha}$.

The embedding of $S$ on $\mathcal G$ imposes non-trivial conditions on $V$ and $A_{\alpha}$. We focus for simplicity on
the case of vanishing Wilson lines, i.e.~$A_\alpha=0$. Firstly, it requires the order of $V$ to be the same
as the order of the generator $\theta$; this is fulfilled if
\begin{equation}
  NV ~\in~ \Lambda\,.
\end{equation}
Secondly, we must impose modular invariance, which, in the absence of Wilson lines, is encoded in~\cite{Bailin:1999nk,Ramos:2008pt,Vaudrevange:2008sm}
\begin{equation}
\label{eq:es}
  N\left( V\cdot V - v\cdot v \right) ~=~ 0\,\,\text{mod 2}\,.
\end{equation}
The easiest way of satisfying this equation is the so-called standard embedding, which consists in taking
vanishing Wilson lines and the shift vector
\begin{equation}
\label{eq:SE}
V ~=~ \left(v^{1},v^{2},v^{3},0,0,0,0,0,0,0,0,0,0,0,0,0 \right)\,.
\end{equation}

It is worth recalling that, for Abelian orbifolds, the resulting 4D gauge group $G$ preserves the rank 16 of the
10D gauge group, \SO{32} in our case. In the standard-embedding scenario, $G$ is further restricted to one of the
following groups~\cite{Ramos:2008pt}:
\begin{equation}
G\in \left\{ \SU{2}^{2}\x\U{1}\x\SO{26},\,\, \SU{2}\x\U{1}^{2}\x\SO{26},\,\, \U{1}^{3}\x\SO{26}  \right\}.
\end{equation}

\subsection{Non-Abelian orbifolds}
\label{sec:NAOrbi}

As a starting point, let us provide here some general useful properties of non-Abelian orbifolds.
First, there are 35 non-Abelian point groups, listed in table~\ref{tab:naoN1}, that lead to 331
inequivalent non-Abelian orbifold geometries compatible with 4D $\mathcal{N}=1$
SUSY~\cite{Fischer:2012qj} when used to compactify a heterotic string theory.
Most of the corresponding point groups can be generated by two non-commuting
group generators, but there are some point groups, such as $(\Z4\x\Z2)\rtimes\Z2$ and $\Delta(216)$,
that require three generators.

\begin{table}[t]
\begin{center}
\begin{tabular}{|c|c|c|c|c|}
\hline
$S_{3}$        &$\Z{3}\x S_{3}$       &$\Z{3}\rtimes\Z{8}$   &$\Z3\x(\Z3\rtimes\Z4)$ &$\Z{3}\x S_{4}$\\
\hline
$D_{4}$        &$\Delta(27)$          &$\SL{2,3}$-I          &$\Z{3}\x A_{4}$        &$\Delta(96)$\\
\hline
$A_{4}$        &$\Z{4}\x S_{3}$       &$\Z{3}\x\SL{2,3}$     &$\Z{6}\x S_{3}$        &$\SL{2,3}\rtimes\Z{4}$\\
\hline
$D_{6}$        &$(\Z6\x\Z2)\rtimes\Z2$&$(\Z4\x\Z4)\rtimes\Z2$&$\Delta(48)$           &$\Sigma(36\phi)$\\
\hline
$\Z8\rtimes\Z2$&$\Z{3}\x D_{4}$   &$\Z3\x((\Z6\x\Z2)\rtimes\Z2)$  &$\GL{2,3}$        &$\Delta(108)$\\
\hline
$QD_{16}$      &$\Z{3}\rtimes Q_{8}$  &$\Delta(216)$         &$\SL{2,3}\rtimes\Z2$   &$\mathrm{PSL}(3,2)$\\
\hline
$(\Z4\x\Z2)\rtimes\Z2$&Frobenius $T_7$&$S_{4}$               &$\Delta(54)$           &$\Sigma(72\phi)$\\
\hline
\end{tabular}
\caption{Non-Abelian point groups compatible with 4D $\mathcal{N}=1$ SUSY. They are 35 different groups giving rise
to 331 geometries~\cite{Fischer:2012qj}.}
 \label{tab:naoN1}
 \end{center}
\end{table}

As a second observation, the Hodge numbers $h^{(1,1)}$ and $h^{(2,1)}$ of the \E8\x\E8 heterotic string compactified on any
supersymmetric non-Abelian orbifold are known~\cite{Fischer:2013qza}. Since these numbers only depend on the topological
properties of the orbifold, they are independent of the kind of string theory that is compactified. That is, their value is the same
for compactifications of the \SO{32} heterotic string. Hence, we already know the number $h^{(1,1)}$ of K\"ahler moduli and
the number $h^{(2,1)}$ of complex-structure moduli of \SO{32} heterotic orbifolds. Furthermore, this information can help to
learn some properties of the massless spectrum due to the relation among moduli and matter representations in the case
of orbifolds with standard embedding~\cite{Dine:1986zy,Dixon:1989fj}. In section~\ref{sec:spectrum}, we will exploit
these results in order to obtain part of the spectrum of non-Abelian heterotic orbifolds.

A third property of non-Abelian orbifolds concerns their fixed points/tori. These fixed loci can be identified
as in Abelian orbifolds by applying eq.~\eqref{eq:fixedpoints1}. However, since all elements of a
conjugacy class yield the same fixed loci~\cite{Fischer:2013qza}, it suffices to determine them
for a representative of the conjugacy class.

We shall discuss in section~\ref{sec:spectrum} a method to embed the orbifold action associated with the
non-Abelian space group $S$ into the gauge degrees of freedom and how to arrive at the matter spectrum
of such orbifolds.

Finally, as it will be discussed in detail in section~\ref{sec:rankr}, the unbroken 4D gauge group $G$ of non-Abelian orbifolds
exhibits rank reduction as a general feature. In particular, in the case of standard embedding the 4D gauge group always has
the form $G=\SO{26}\x \widetilde{G}$, where $\widetilde{G}$ is such that $\operatorname{rank}(\widetilde{G})<3$.

%% file: rankreduction.tex
\section{Top-down rank reduction}
\label{sec:rankr}
One of the most important features of the non-Abelian scenario is the rank reduction of the original gauge group.
In the bottom-up approach it is possible to reduce the rank of a gauge group through
non-Abelian twists~\cite{Hebecker:2003}. Such twists appear naturally in the top-down approach based on
non-Abelian orbifold compactifications of heterotic strings. Hence, it should be possible
to provide rank reduction from the top-down approach in this context. Let us elaborate on this.

As mentioned earlier, in the case of Abelian point groups, $P$ is not only a subgroup of \SO6, but it is a subgroup of the \SO6 Cartan
subalgebra $H_{\SO6}$. Hence, with a single basis for the spatial compact dimensions and a choice of the Cartan basis, it is possible to express
every single element $h$ of $P$ as the following exponential map
\begin{equation} 
  h~=~\exp\left[\frac{2\pi\I}{n_{h}}\left( \alpha_{1} J_{3,4} + \alpha_{2} J_{5,6} + \alpha_{3} J_{7,8} \right)  \right]\,,
\end{equation}
where $n_{h}$ is the order of the element $h$, $\alpha_{j}\in\Z{}$ are different for each $h$ and $H_{\SO6}=\{J_{3,4}, J_{5,6},J_{7,8}\}$
is the usual Cartan basis of \SO6 with $J_{i,j}$ the generator of rotations in the plane $X^{i}-X^{j}$. 

In the non-Abelian case, it turns out that $P$ is not entirely contained in the \SO6 Cartan subalgebra.
This is the origin of the rank reduction of the gauge group.
To see this explicitly, consider that for a non-Abelian $P$, there is at least one $g\in P$, such that
\begin{equation}
  g~\neq~ \exp\left[ \frac{2\pi\I}{n_g}\left(\alpha_{1}J_{3,4}+\alpha_{2}J_{5,6}+\alpha_{3}J_{7,8}  \right) \right]\,,
\end{equation}
because, otherwise, it would commute with all other elements $h$ as given above. This implies that
such non-commuting point-group element can take the form
\begin{equation}
  g ~=~ \exp\left[ \frac{2\pi\I}{n_g}\left(c_{1} J_{1} +c_{2} J_{2} +c_{3} J_{3}  \right) \right]\,,
\end{equation}
for some choice $\{J_{1}, J_{2},J_{3}\}$ of \SO6 generators and $c_j\in\Z{}$,
such that at least one $J_{i}\not\in H_{\SO6}$, i.e.\ it does not commute with the elements of $H_{\SO6}$,
explaining why $[g,h]\neq0$.

A crucial step to arrive at rank reduction is that the orbifold is embedded into the gauge degrees of freedom.
In the case of standard embedding that we study in this work, the orbifold action is identical on the \SO6 subgroup
of the \SO{32} gauge group. To be orbifold invariant, a Cartan generator must be compatible with all
point-group elements. Since the generator $g$ is not expressed in terms of the \SO6 Cartan subalgebra,
it has a non-trivial action of at least a Cartan generator of the \SO{32} gauge group, which is hence projected out.
Thus, \SO{32} Cartan generators are projected out by the orbifold action.

Since in the case of \SO{32} heterotic orbifolds with standard embedding the orbifold acts on the \SO6 subgroup
of \SO{32}, it naturally leaves invariant the \SO{26} gauge subgroup of rank 13. We expect thus that the non-Abelian structure
of $P$ leads to the projection for some (or even all three) of the Cartan generators of the \SO6 gauge subgroup,
reducing the rank by up to three units (i.e.\ from 16 to up to 13). As we show in section~\ref{sec:results},
in the $(\Z{4}\x\Z{2})\rtimes\Z{2}$ orbifold, no \SO{6} Cartan element survives the orbifold projection,
yielding the 4D gauge group $G=\SO{26}$.

%% file: spectrum.tex
\section{Massless spectrum of non-Abelian orbifolds}
\label{sec:spectrum}

To date, there has been only one effort to obtain the full massless matter spectrum of heterotic orbifolds,
focusing on the compactification of the $\E8\x\E8$ heterotic string on an orbifold with point group $P=S_{3}$~\cite{Konopka:2012gy}.
In this work, we aim to extend that pursuit.

Let us first recall that the massless matter spectrum of heterotic orbifolds is built by two types of strings:
untwisted and twisted. Untwisted strings are those that are closed in the uncompactified heterotic string
and are left invariant by the orbifold action. Untwisted states are collected together in the untwisted sector $U$.
The unbroken gauge group $G$, geometric moduli and some matter
states are examples of effective states arising from untwisted strings.
Twisted strings are only closed due to the action of the orbifold and are attached to the orbifold singularities,
where the orbifold action is encoded by the constructing elements of each fixed locus (see section~\ref{sec:AOrbi}).
In non-Abelian orbifolds, they are collected in different twisted sectors $T_{[\theta]}$ associated with each
conjugacy class $[\theta]$ of the point group. As mentioned before, for each $[\theta]\in P$ there is a number
of constructing elements corresponding to the fixed loci of the sector.

As a first approach, since the number of moduli of non-Abelian orbifolds has been computed~\cite{Fischer:2013qza}
and a connection among moduli and massless matter is known~\cite{Dine:1986zy,Dixon:1989fj}, we can determine the
non-trivial matter representations under the \SO{26} factor of the unbroken 4D gauge group. In \E8\x\E8 heterotic
orbifolds with standard embedding, each K\"ahler modulus yields is associated with a fundamental $\rep{27}$ representation of
the unbroken \E6 gauge factor whereas each complex-structure modulus is related to an anti-fundamental $\crep{27}$.
Consequently, there are $h^{(1,1)}$ $\rep{27}$ and $h^{(2,1)}$ $\crep{27}$ of \E6 in the massless spectrum.
It is straightforward to generalize this result to \SO{32} heterotic orbifold. Since the fundamental
$\rep{26}$ representation of the unbroken \SO{26} gauge factor is real, one finds that both types of moduli
lead to fundamental representations (plus singlets) in the massless spectrum. That is, one finds
$h^{(1,1)}+h^{(2,1)}$ fundamental $\rep{26}$ representations of \SO{26} of every non-Abelian orbifold.

Based on these observations and the results of ref.~\cite{Fischer:2013qza}, we have determined the
non-trivial representations of the \SO{26} factor of the unbroken 4D gauge group for all non-Abelian heterotic
orbifolds. We list our results in table~\ref{tab:n26} of Appendix~\ref{app:26s}. In table~\ref{tab:n26summary},
we present an excerpt thereof containing the information of three sample cases free of roto-translations,
which shall be studied in further detail below.

\begin{table}[t]
\begin{center}
\resizebox{\textwidth}{!}{
\begin{tabular}{|c|c|c|c|}
\hline
Point group & \multirow{2}{*}{\Z{} class -- affine class} & Generators of $P$ & \multirow{2}{*}{\#$\rep{26}$} \\
\cline{3-3}
$P$ &  & Number of \boldmath$26\,$\unboldmath from $U$ and $T$ sectors &  \\
\hline
\multirow{2}{*}{$S_{3}$} & \multirow{2}{*}{$1-1$} & $(\vartheta,0),\,(\omega,0)$ & \multirow{2}{*}{30}  \\
\cline{3-3}
&   & $(4)U + (8)T_{[\vartheta]}+ (18)T_{[\omega]}$ & \\
\hline
\multirow{2}{*}{$D_{4}$} & \multirow{2}{*}{$5-1$} & $(\vartheta,0),\,(\omega,0)$ & \multirow{2}{*}{42}  \\
\cline{3-3}
&   & $(4)U + (4)T_{[\vartheta]}+ (16)T_{[\omega]}+ (8)T_{[\vartheta\omega]}+ (10)T_{[(\vartheta\omega)^{2}]}$ &  \\
\hline
\multirow{3}{*}{$(\Z{4}\x\Z{2})\rtimes\Z{2}$} & \multirow{3}{*}{$2-1$} & $(\vartheta,0),\,(\omega,0),\,(\rho,0)$ & \multirow{3}{*}{38} \\
\cline{3-3}
&  & $(3)U + (4)T_{[\vartheta]}+ (4)T_{[\omega]}+ (10)T_{[\rho]}+ (2)T_{[\vartheta\omega]}$ & \\
&   & $+ (2)T_{[\vartheta\rho]}+ (2)T_{[\omega\rho]}+ (6)T_{[\vartheta\omega\rho^{3}]}+
(5)T_{[\rho^{3}]}$ & \\
\hline
\end{tabular}
}
\caption{Fundamental representations of $\SO{26}$ from the untwisted and twisted sectors of some $S_{3}$, $D_{4}$ and $(\Z{4}\x\Z{2})\rtimes\Z{2}$
orbifolds. In our notation, $(n) U$ means that there are $n$ $\rep{26}$'s in the untwisted sector, while $(m) T_{[\theta]}$ denote the presence of $m$ $\rep{26}$'s
in the twisted sector associated with the conjugacy class $[\theta]$ of $P$.}
 \label{tab:n26summary}
 \end{center}
\end{table}

On the other hand, it is well known how to compute the massless spectrum of Abelian heterotic
orbifolds~\cite{Bailin:1999nk,Ramos-Sanchez:2008nwx,Vaudrevange:2008sm}.
The techniques used in that case are our starting point to figure out the detailed spectrum of non-Abelian orbifolds.
We shall address in the following the possible generalization of the Abelian techniques.

Let us focus on orbifolds with no roto-translations.
Given a particular model with point group $P$, in order to compute the spectrum the first step is to determine the constructing elements of
the group, i.e.\ the representative elements of the non-equivalent conjugacy classes. Once this is achieved, we have to determine the
fixed points for each rotational part of each constructing element $\theta\in P$ by solving an analogue equation to eq.~\eqref{eq:fixedpoints1}
\begin{equation} \label{eq:fixedpoints}
Z_{f} ~=~ \left( \Id -\theta \right)^{-1} n_{\alpha}e_{\alpha}\,,\qquad n_\alpha\in\Z{}\,,
\end{equation}
where $e_{\alpha}$ are the basis vectors of the toroidal lattice $\Gamma$.
This process is identical for Abelian and non-Abelian orbifolds.

The next task, and the first non-trivial one, is to obtain the twist and shift vectors for each constructing element. For this, we have to embed 
the geometric group $P$ into the gauge degrees of freedom. This differs significantly from the procedure in the Abelian case due to the
non-Abelian nature of our problem. In the Abelian instance, the group $P$ is generated by the Cartan subalgebra, becoming an Abelian subgroup by definition,
which is realized in both the geometric and the gauge degrees of freedom. This cannot happen in the non-Abelian scenario. This fact forces us to work with a different
Cartan basis for each twisted sector, and makes clear the need of posing certain changes to the existing methods for calculating the 4D gauge group,
the spectrum and so on.

Focusing on the standard embedding, the embedding of $P$ in the gauge degrees of freedom,
$P\hookrightarrow \SO{32}$, takes the following structure
\begin{equation}\label{eq:PgaugeEm}
P\hookrightarrow \SO{6}\subset\SO{6}\times\SO{26}\subset\SO{32}.
\end{equation}
To achieve~\eqref{eq:PgaugeEm} it is enough to express every conjugacy class of $P$ as an exponential map of a linear combination 
of the generators of \SO{6}. For this purpose, we develop an algorithm to achieve block diagonalization as detailed in
Appendix~\ref{app:BDdetails}. With this treatment, we are able to express each conjugacy class as an exponential of a linear 
combination of (at most) three \SO{6} generators, e.g.
\begin{equation}\label{eq:ccExp}
g ~=~ \exp\left[ \frac{2\pi\I}{n_g}\left( \alpha_{1}J_{1} + \alpha_{2}J_{2} + \alpha_{3}J_{3}  \right) \right],
\end{equation}
for some (roto-translation-free) generator $g\in S$ of order $n_g$.

Once we obtain eq.~\eqref{eq:ccExp} for each conjugacy class, we continue to solve
\begin{equation} 
\left[ \sum_{j=1}^{15} \beta_{j}J_{j} ~,~  \alpha_{1}J_{1} + \alpha_{2}J_{2} + \alpha_{3}J_{3} \right] ~=~ 0\,,
\end{equation}
simultaneously for every conjugacy class. The solutions are the generators of the remaining subgroup $\widetilde{G}$ of \SO{6} after 
the compactification, so our 4D gauge group in the standard embedding scenario is 
\begin{equation}
G ~=~ \widetilde{G}\times\SO{26}\,.
\end{equation} 
Note that the $\widetilde{G}$ group can be determined by inspecting the relation among all generators, which
requires identifying the orbifold-invariant, massless charged gauge bosons and their algebra. So, we must
determine the properties of the massless matter spectrum.

From section~\ref{sec:SO32}, recall that the original 10D massless states have the form $\ket{p}_{L}\otimes \ket{q}_{R}$,
subject to the masslessness condition from eq.~\eqref{eq:masslessSO32}. To obtain the 4D states, we have to compactify the six remaining 
spatial dimensions on an orbifold. This compactification imposes additional boundary conditions, given by
\begin{equation} \label{eq:tsboundaryc}
  Z^{a}(\tau,\sigma+\pi) ~=~ (g Z)^{a}(\tau,\sigma),\quad a\in\left\{1,2,3 \right\}\,,
\end{equation}
with $g\in S$ and $(\tau,\sigma)$ the usual parameters of the string world-sheet. Thus, there are new extra closed strings over 
the compact dimensions. They are associated with the element $g$ in turn, and with its fixed points in the orbifold.
The corresponding matter spectrum comprise the twisted sectors of our theory.
As mentioned above, there are as many twisted sectors as inequivalent conjugacy classes in the point group.
It is important to notice that the only element in $S$ that does not modify the boundary conditions is
the identity $e$, which builds the untwisted sector $U$.

\subsection[Matter states from the untwisted sector U]{\boldmath Matter states from the untwisted sector $U$\unboldmath}
\label{ssec:untwistedspectra}

The 4D massless states can be built as usual by tensoring together a right-moving mode $\ket{q}_{R}$
and a left-moving mode $\ket{p}_{L}$. In the untwisted sector, the boundary conditions are not modified
and the states must satisfy the same masslessness conditions~\eqref{eq:masslessSO32} as the 10D states.
Now, not every state $\ket{p}_{L}\otimes \ket{q}_{R}$ is a physical state. To note this, consider their transformation under the 
action of an element $g\in S$ of order $n_g$
\begin{equation} \label{eq:orbitransform}
\ket{p}_{L}\otimes \ket{q}_{R}~\xmapsto{~g~}~ \exp\left[ 2\pi \I \left( p\cdot V_{g} - q\cdot v_{g} \right) \right]
\left(  \ket{p}_{L}\otimes \ket{q}_{R} \right), 
\end{equation}
where the twist vector is defined as $v_{g}=\frac{1}{n_g}(0,\alpha_1,\alpha_2,\alpha_3)$, with $\alpha_{i}$ the coefficients 
from eq.~\eqref{eq:ccExp}. Further, the shift vector takes the form $V_{g}=\frac{1}{n_g}(\alpha_1,\alpha_2,\alpha_3,0^{13})$
for the standard embedding, cf.\ eq.~\eqref{eq:SE}, where the power denotes entry repetition.
Hence, a massless state is invariant under the conjugacy class represented by $g$ if
\begin{equation} \label{eq:Uorbinvariant}
 p\cdot V_{g} - q\cdot v_{g} ~=~ 0\bmod 1\,.
\end{equation}
In addition, to be fully orbifold invariant, the states must satisfy a similar invariance condition
for all elements in the centralizer of $g$. In the absence of roto-translations, $g$ can be regarded
as a point-group element. In this scenario, valid elements of the centralizer can be found in $\mathcal{C}_{P}(g)$,
which is defined by
\begin{equation}
\mathcal{C}_{P}(g) ~=~ \left\{ h\in P ~|~ [g,h]=0 \right\}\,.
\end{equation} 

Note that in the case of Abelian orbifolds, the centralizer $\mathcal{C}_{P}(g)$ of every element $g$ is Abelian. This is not
always true in the non-Abelian case. From this we conjecture that in order to compute the spectrum we have to apply different 
procedures depending on whether $\mathcal{C}_{P}(g)$ is Abelian or not. For the untwisted sector, $\mathcal{C}_{P}(e)=P$
is always non-Abelian, while this needs not be the case for the twisted sectors. Then, a different treatment has to be implemented 
depending on the sector of interest. 

From eq.~\eqref{eq:Uorbinvariant}, we see that orbifold invariance requires to find the states that are invariant under
the centralizer of the identity element $\mathcal{C}_{P}(e)$, i.e.\ under the full group. This is challenging since each conjugacy class has its twist 
vector written in a different Cartan basis. So, we have to consider eq.~\eqref{eq:Uorbinvariant} in every chosen basis and solve it simultaneously for all bases.
To remark the difficulty that this computation represents, recall that the states $\ket{p}_{L}\otimes \ket{q}_{R}$ in the untwisted 
sector are built with the weights $q$ describing the $\rep8_s$ and $\rep8_v$ representations of \SO{8} and $p$ a simple root of \SO{32}. 
As a consequence, its explicit form in coordinates depends on the choice of the Cartan basis from which the simple root system is derived. 

Taking these ideas into account, our method relies on computing the simple-root system for the choice of the Cartan basis, and then building
a consistent ordering for the simple roots as follows. Regarding the $q$'s in the spectrum, take two different choices $H$ and $H'$ for the Cartan basis of \SO{32},
with elements $H_{i}$ and $H'_{i}$, respectively, with $i\in \{1,2,\ldots,16 \}$, and compute the 32 simple roots for each basis, 
$R=\{R_{1},R_{2},\ldots,R_{32}\}$ and $R'=\{R'_{1},R'_{2},\ldots,R'_{32}\}$, correspondingly. Now, the problem we have is that
there is no transformation $Q$, such that
\begin{equation}
  R \xrightarrow{Q} R'\, .
\end{equation}
Hence, in order to work simultaneously with $R$ and $R'$, we have to identify their elements in a clever way. Note that, since $H$ and $H'$
are ordered bases, the physical roll that the element $H_{i}$ plays is equivalent to the one that $H'_{i}$ does. From this fact, we impose the
identification $H_{i}\sim H'_{i}$, for each $i$. So we can speak about the $n$-th Cartan generator without giving an explicit form of it. Next,
we consider the simple roots for each basis, $R$ and $R'$. Suppose that $R_\ell$ acts as a raising operator for $H_{1}$, while $R_{m}$ behaves
as its lowering operator for some $\ell$ and $m$. The same behavior is exhibited by two elements of $R'$: assume that $R'_\ell$ and $R'_{m}$ are
the raising and lowering operators for $H'_{1}$. Since we have identified $H_{1}\sim H'_{1}$ due to its physical roll, the simple roots will inherit this
identification as $R_\ell\sim R'_\ell$ and $R_{m}\sim R'_{m}$. These relations can be extended in the same way to every element in $R$ and $R'$, 
by direct computation of the rising and lowering operators of $H_{2},\ldots, H_{16},$ $H'_{2},\ldots,H'_{15}$ and $H'_{16}$. 
Naturally, an analogous procedure has to be applied to the case of \SO{6} for the $q$'s in the spectrum. 

In this way, we have built a ``bijection'' between the elements of $R$ and $R'$, and we can identify them in a basis-independent way.
This step is crucial for our method. By using it, we ensure the construction of labels that are independent of the basis choice for each simple root
and we can proceed with the orbifold projection keeping only those states that are invariant under each twist vector on its respective basis. 

Now that we have developed a method to solve eq.~\eqref{eq:Uorbinvariant}, we can proceed to study its solutions.
As has been addressed~\cite{Ramos:2008pt,Vaudrevange:2008sm}, the nature of the states in the untwisted sector fully depend on whether the right-moving
modes satisfy $q\cdot v_{g}=0$ or $q\cdot v_{g}\neq0$. Those states in the first category constitute the gauge sector of the spectrum, i.e.\ the unbroken
gauge fields for the 4D gauge group. States under the second non-trivial condition correspond to massless matter representations.

\subsection[The twisted sectors Tg]{\boldmath The twisted sectors $T_{[g]}$ \unboldmath}
\label{ssec:twistedspectra}

Consider a constructing element $g$. The states in its twisted sector are constituted as $|q\rangle_{R} \otimes |p\rangle_{L}$, with $|q\rangle_{R}$
either in the spinor or fermion weight lattice of \SO{8}, and $p\in\Lambda$. Due to eq.~\eqref{eq:tsboundaryc}, $q$ and $p$ are solutions of the 
twisted masslessness conditions, given by
\begin{subequations} \label{eq:twsmassless}
\begin{align}
\frac{q_{sh}\cdot q_{sh}}{2} -\frac{1}{2} + \delta_{g}=0, & \quad q_{sh}=q+v_{g},\\
 \frac{p_{sh}\cdot p_{sh} }{2} -1 + \tilde{N} + \delta_{g}=0, & \quad p_{sh}=p+V_{g},
 \end{align}
\end{subequations}
with the zero-point energy shift $\delta_{g}$ defined as 
\begin{equation}
\delta_{g} ~=~ \frac{1}{2} \sum_{i=0}^{3} \eta_{i}\left( 1-\eta_{i} \right)\,,
\end{equation}
where $\eta_{i}=(v_{g})_{i}\,\bmod 1$.

Again, not every solution to eq.~\eqref{eq:twsmassless} is a physical state. The action of an element $h\in S$ over any state of the twisted sector
associated with the constructing element $g$ reads
\begin{equation} \label{eq:twsorbitransform}
\ket{p_{sh}}_{L}\otimes \ket{q_{sh}}_{R}\rightarrow \exp\left[ 2\pi \I \left( p_{sh}\cdot V_{h} - Q_{sh}\cdot v_{h} \right) \right]\exp[\Phi_{vac}]
\ket{p_{sh}}_{L}\otimes \ket{q_{sh}}_{R}, 
\end{equation}
with $\Phi_{vac}$ the vacuum phase, defined by
\begin{equation}
\Phi_{vac} ~:=~ V_{h}\cdot V_{g} - v_{h}\cdot v_{g}\,,
\end{equation}
and $Q_{sh}^{i}$, the $R-$charges as often referred to in the literature~\cite{Araki:2007ss,CaboBizet:2013gns,Nilles:2013lda}, written as
\begin{equation}
Q_{sh}^{i} ~:=~ q_{sh}^{i}-\tilde{N}^{i}+\tilde{N}^{*i}\,.
\end{equation}
Here, $\tilde{N}^{i}\,(\tilde{N}^{*i})$ count the number of holomorphic(antiholomorphic) oscillators. Now, for standard embedding, the vacuum phase is
always trivial. Consequently, the state is invariant if 
\begin{equation}\label{eq:twsorbinvariant}
p_{sh}\cdot V_{h} - Q_{sh}\cdot v_{h}  ~=~ 0\bmod 1
\end{equation}
for every element $h$ in the centralizer of $g$, denoted as $\mathcal{C}_{P}(g)$. 

Now, we have two cases: the centralizer $\mathcal{C}_{P}(g)$ is either a non-Abelian or an Abelian subgroup of $P$. In the first case, we have to deal
with the same issues as in the untwisted sector and the procedure described previously must be replicated. In the second case, we are in the same case as
in Abelian orbifolds and we can proceed directly, since every element of the centralizer can be treated in the same Cartan basis.

For the sake of clarity, let us discuss in detail the implementation of our method in the case of the non-Abelian orbifold
with the symmetric group $S_{3}$ as point group.

\subsection[Example: S3 orbifold]{\boldmath Example: $S_{3}$ orbifold\unboldmath }

We work with the following presentation of the $S_{3}$ group 
\begin{equation}\label{eq:S3pre} 
S_{3}~=~\langle \vartheta,\,\omega\,\, | \,\,\vartheta^{2}=\omega^{3}=e \rangle\,.
\end{equation}
This group has three conjugacy classes:
\begin{equation}\label{eq:S3cc} 
[e]=\{e\},\quad [\vartheta]=\{\vartheta,\,\vartheta\omega,\,\vartheta\omega^{2}\},\quad [\omega]=\{\omega,\,\omega^{2}\}.
\end{equation}
 
In order to fully specify our orbifold, we have to provide a representation of the point group generators and the basis for the torus lattice.
To represent the generators,
we use the same matrices appearing in~\cite{Konopka:2012gy}. So, $\vartheta$ and $\omega$ have the form
\begin{equation}\label{eq:S3generators}
\vartheta=\begin{pmatrix}
1 & -1 & 0 & 0 & 0 & 0\\
0 & -1 & 0 & 0 & 0 & 0\\
0 & 0 & 1 & -1 & 0 & 0\\
0 & 0 & 0 & -1 & 0 & 0\\
0 & 0 & 0 & 0 & -1 & 0\\
0 & 0 & 0 & 0 & 0 & -1
\end{pmatrix},\qquad
\omega=\begin{pmatrix}
0 & -1 & 0 & 0 & 0 & 0\\
1 & -1 & 0 & 0 & 0 & 0\\
0 & 0 & -1 & 1 & 0 & 0\\
0 & 0 & -1 & 0 & 0 & 0\\
0 & 0 & 0 & 0 & 1 & 0\\
0 & 0 & 0 & 0 & 0 & 1
\end{pmatrix},
\end{equation}
On the other hand, the lattice that defines our torus is generated by the vectors $e'_{i}\in\mathbbm{R}^{6}$, given by
\begin{equation}\label{eq:S3latticebasis}
\begin{split}
e'_{3}&=\left(1,0,0,0,0,0\right)^{\operatorname{T}}, \qquad e'_{4}=\left(-\frac{1}{2},\frac{\sqrt{3}}{2},0,0,0,0\right)^{\operatorname{T}},\\
e'_{5}&=\left(0,0,1,0,0,0\right)^{\operatorname{T}}, \qquad e'_{6}=\left(0,0,-\frac{1}{2},\frac{\sqrt{3}}{2},0,0\right)^{\operatorname{T}},\\
e'_{7}&=\left(0,0,0,0,1,0\right)^{\operatorname{T}}, \qquad e'_{8}=\left(0,0,0,0,-\frac{1}{2},\frac{\sqrt{3}}{2}\right)^{\operatorname{T}}.
\end{split}
\end{equation} 
The orbifold specified by eqs.~\eqref{eq:S3generators} and~\eqref{eq:S3latticebasis} has 13 fixed points, 4 of them invariant under $[\vartheta]$ and
9 of them fixed by $[\omega]$. 

We notice that, if we switch to the canonical Euclidean basis
\begin{equation}\label{eq:S3latticecanonical}
\begin{split}
e'_{3}&=\left(1,0,0,0,0,0\right)^{\operatorname{T}}, \qquad e'_{4}=\left(0,1,0,0,0,0\right)^{\operatorname{T}},\\ 
e'_{5}&=\left(0,0,1,0,0,0\right)^{\operatorname{T}}, \qquad e'_{6}=\left(0,0,0,1,0,0\right)^{\operatorname{T}},\\ 
e'_{7}&=\left(0,0,0,0,1,0\right)^{\operatorname{T}}, \qquad e'_{8}=\left(0,0,0,0,0,1\right)^{\operatorname{T}}, 
\end{split}
\end{equation} 
via the transformation
\begin{equation}
\vartheta\to Q\vartheta Q^{-1},\qquad \omega\to Q\omega Q^{-1},
\end{equation}
with $Q$ given by 
\begin{equation}\label{eq:S3Q}
Q=
\left(
\begin{array}{cccccc}
 1 & -\frac{1}{2} & 0 & 0 & 0 & 0 \\
 0 & \frac{\sqrt{3}}{2} & 0 & 0 & 0 & 0 \\
 0 & 0 & 1 & -\frac{1}{2} & 0 & 0 \\
 0 & 0 & 0 & \frac{\sqrt{3}}{2} & 0 & 0 \\
 0 & 0 & 0 & 0 & 1 & -\frac{1}{2} \\
 0 & 0 & 0 & 0 & 0 & \frac{\sqrt{3}}{2} \\
\end{array}
\right),
\end{equation} 
we obtain a more useful block-diagonal form for the generators $\vartheta$ and $\omega$,
\begin{equation}\label{eq:S3blockgen}
\vartheta=\begin{pmatrix}
1 & 0 & 0 & 0 & 0 & 0\\
0 & -1 & 0 & 0 & 0 & 0\\
0 & 0 & 1 & 0 & 0 & 0\\
0 & 0 & 0 & -1 & 0 & 0\\
0 & 0 & 0 & 0 & -1 & 0\\
0 & 0 & 0 & 0 & 0 & -1
\end{pmatrix},\qquad
\omega=\begin{pmatrix}
-\frac{1}{2} & -\frac{\sqrt{3}}{2} & 0 & 0 & 0 & 0\\
\frac{\sqrt{3}}{2} & -\frac{1}{2} & 0 & 0 & 0 & 0\\
0 & 0 & -\frac{1}{2} & \frac{\sqrt{3}}{2} & 0 & 0\\
0 & 0 & -\frac{\sqrt{3}}{2} & -\frac{1}{2} & 0 & 0\\
0 & 0 & 0 & 0 & 1 & 0\\
0 & 0 & 0 & 0 & 0 & 1
\end{pmatrix}.
\end{equation} 
It is crucial that the $S_3$ group invariants remain invariant under the change of basis governed by $Q$; otherwise, $Q$ would
spoil orbifold invariance and would wrongly lead to projecting out relevant physical states. We explicitly verify in Appendix~\ref{app:GT}
the consistency of orbifold invariance under the basis change. This result can be generalized to all orbifold geometries.

From eq.~\eqref{eq:S3blockgen}, we can express our generators as exponential maps of the \SO{6} generators 
\begin{equation}\label{eq:S3expgen}
\vartheta=\operatorname{exp}\left[ \frac{2\pi\I}{2}\left( J_{4,6} - J_{7,8} \right) \right],\qquad
\omega=\operatorname{exp}\left[ \frac{2\pi\I}{3}\left( J_{3,4} - J_{5,6} \right) \right].
\end{equation} 
Here, we observe that these twists are given in different Cartan bases, as we anticipated.
Explicitly, the choices for our two Cartan bases for \SO{6} are
\begin{subequations}\label{eq:S3CartanBso6}
\begin{align}
H_{\vartheta}^{\SO{6}}&=\left\{ J_{3,5},J_{4,6},J_{7,8} \right\},\\ 
H_{\omega}^{\SO{6}}&=\left\{ J_{3,4},J_{5,6},J_{7,8} \right\}.
\end{align} 
\end{subequations}
Using these bases, we write the twists vectors in their corresponding basis as
\begin{equation} \label{eq:S3twist}
v_{\vartheta}=\left( 0,0,\frac{1}{2},-\frac{1}{2} \right),\qquad
v_{\omega}=\left( 0,\frac{1}{3},-\frac{1}{3},0 \right).
\end{equation}
Embedding the choices~\eqref{eq:S3CartanBso6} for the \SO{6} Cartan bases into \SO{32} by demanding standard embedding leads to the \SO{32}
Cartan bases given by
\begin{subequations}\label{eq:S3CartanBso32}
\begin{align}
H_{\vartheta}^{\SO{32}}&=\left\{ J_{1,2},J_{3,5},J_{4,6},J_{7,8},J_{9,10},J_{11,12},J_{13,14},\ldots,J_{31,32} \right\},\\
H_{\omega}^{\SO{32}}&=\left\{ J_{1,2},J_{3,4},J_{5,6},J_{7,8},J_{9,10},J_{11,12},J_{13,14},\ldots,J_{31,32} \right\}.
\end{align} 
\end{subequations}
From eqs.~\eqref{eq:S3CartanBso6} and~\eqref{eq:S3CartanBso32} it follows that the shift vectors for each sector are
\begin{equation} \label{eq:S3shift}
V_{\vartheta}=\left( 0,\frac{1}{2},-\frac{1}{2},0^{13}  \right),\qquad 
V_{\omega}=\left( \frac{1}{3},-\frac{1}{3},0,0^{13} \right). 
\end{equation}

We can now compute the 4D gauge group. We must identify the linear combinations of the generators of \SO{6}
that commute simultaneously with the arguments of the exponentials in eq.~\eqref{eq:S3expgen}.
It turns out that there are only two of them, and they are given by
\begin{equation}\label{eq:S3ggG}
T_{1}~=~J_{3,5}-J_{4,6} \qquad\text{and}\qquad T_{2}~=~J_{7,8}\,.
\end{equation}
Now, as will be shown, in the spectrum of the untwisted sector we find only 13 Cartan generators and 312 roots. From this, we conclude that
both $T_{1}$ and $T_{2}$ are generators of \U{1} symmetries. Hence, we have the following gauge symmetry breaking
\begin{equation}\label{eq:S3gg}
\SO{32} ~\to~ \U{1}\x\U{1}\x\SO{26} ~=~ G_{S_{3}}\,.
\end{equation} 
From eqs.~\eqref{eq:S3ggG} and~\eqref{eq:S3gg}, it is clear that we have reduced the rank of the gauge group. Explicitly, $\operatorname{rank}(\SO{32})=16$,
while $\operatorname{rank}(G_{S_{3}})=15$. This is the minimal rank reduction that can be obtained in the non-Abelian scenario. 

Now, to obtain the matter spectrum, we determine the simple roots of \SO{32} for both $H_{\vartheta}^{\SO{32}}$ and $H_{\omega}^{\SO{32}}$.
The simple roots are 
\begin{subequations}
\begin{align}
R_{\vartheta}^{\SO{32}}={}&  \left\{  (1,-1,0^{14}),(0,1,-1,0^{13}),(0,0,1,-1,0^{12}),(0^{3},1,-1,0^{11}),(0^{4},1,-1,0^{10}),\right.\\
&\left.(0^{5},1,-1,0^{9}),(0^{6},1,-1,0^{8}),(0^{7},1,-1,0^{7}),(0^{8},1,-1,0^{6}),(0^{9},1,-1,0^{5}),\right.\\
&\left.(0^{10},1,-1,0^{4}),(0^{11},1,-1,0^{3}),(0^{12},1,-1,0^{2}),(0^{13},1,-1,0),(0^{14},1,-1),(0^{14},1,1) \right\}, 
\end{align}

\begin{align}
R_{\omega}^{\SO{32}}={}&  \left\{  (1,-1,0^{14}),(0,1,-1,0^{13}),(0,0,1,-1,0^{12}),(0^{3},1,-1,0^{11}),(0^{4},1,-1,0^{10}),\right.\\
&\left.(0^{5},1,-1,0^{9}),(0^{6},1,-1,0^{8}),(0^{7},1,-1,0^{7}),(0^{8},1,-1,0^{6}),(0^{9},1,-1,0^{5}),\right.\\
&\left.(0^{10},1,-1,0^{4}),(0^{11},1,-1,0^{3}),(0^{12},1,-1,0^{2}),(0^{13},1,-1,0),(0^{14},1,-1),(0^{14},1,1) \right\}.
\end{align}
\end{subequations}

On the other hand, for \SO{6} we have 
\begin{subequations}
\begin{align}
R_{\vartheta}^{\SO{6}}=&\left\{  (1,-1,0^{4}),(0,1,-1,0^{3}),(0^{2},1,-1,0^{2}),(0^{3},1,-1,0),(0^{4},1,-1),(0^{4},1,1)   \right\},\\ 
R_{\omega}^{\SO{6}}=&\left\{ (1,-1,0^{4}),(0,1,-1,0^{3}),(0^{2},1,-1,0^{2}),(0^{3},1,-1,0),(0^{4},1,-1),(0^{4},1,1) \right\}.
\end{align} 
\end{subequations}

Even though $R_{\omega}$ and $R_{\vartheta}$ look alike for both \SO{32} and \SO{6}, they should not be taken as identical objects. Due to 
eqs.~\eqref{eq:S3CartanBso6} and~\eqref{eq:S3CartanBso32}, the elements of $R_{\omega}$ and $R_{\vartheta}$ are linear combinations of different bases.
Hence, even if the coefficients of the linear combinations are exactly the same, they cannot be manipulated simultaneously.

With this in hand, we solve eq.~\eqref{eq:Uorbinvariant} and find the states in the untwisted sector. 
For the twisted sectors, we have to calculate the centralizers first. Using GAP~\cite{GAP4}, we find that
\begin{equation}
\mathcal{C}_{S_{3}}(\vartheta)~=~\Z{2}\qquad\text{and}\qquad\mathcal{C}_{S_{3}}(\omega)~=~\Z{3}\,.
\end{equation} 
Therefore, we are back at the Abelian case: the standard Abelian methods can produce the spectrum of both twisted sectors.
The results are presented in table~\ref{tab:S3spectrum}.
 
It is important to notice that the number of $\rep{26}$ is in agreement with the results obtained {\it a priori}
from topological properties as displayed in table~\ref{tab:n26summary}.

\begin{table}[t]
\begin{center}
\begin{tabular}{|c|c|c|c|}
\hline 
$\SO{26}\times\U{1}\times\U{1}$ irrep. & $U$ sector & $T_{[\vartheta]}$ sector & $T_{[\omega]}$ sector  \\ 
\hline 
$\rep{26}_{(1,0)}$ & 1 & 0 & 9 \\ 
\hline 
$\rep{26}_{(-1,0)}$ & 1 & 0 & 9 \\ 
\hline 
$\rep{26}_{(0,1)}$ & 1 & 0 & 0 \\ 
\hline 
$\rep{26}_{(0,-1)}$ & 1 & 0 & 0 \\ 
\hline 
$\rep{26}_{(-1/2,\,-1/2)}$ & 0 & 4 & 0 \\ 
\hline 
$\rep{26}_{(1/2,\,1/2)}$ & 0 & 4 & 0 \\ 
\hline 
$\rep{1}_{(0,0)}$ & 3 & 0 & 18  \\ 
\hline 
$\rep{1}_{(1/2,\,-1/2)}$ & 0 & 8 & 0  \\ 
\hline 
$\rep{1}_{(-1/2,\,1/2)}$ & 0 & 8 & 0  \\ 
\hline 
$\rep{1}_{(-3/2,\,-1/2)}$ & 0 & 4 & 0  \\ 
\hline 
$\rep{1}_{(3/2,\,1/2)}$ & 0 & 4 & 0  \\ 
\hline
$\rep{1}_{(1,1)}$  & 1 & 0 & 9 \\ 
\hline
 $\rep{1}_{(1,-1)}$ & 1 & 0 & 9 \\ 
 \hline  
$\rep{1}_{(-1,1)}$ & 1 & 0 & 9 \\ 
\hline
$\rep{1}_{(-1,-1)}$ & 1 & 0 & 9 \\ 
\hline 

\end{tabular}
\caption{Massless matter spectrum of the $S_3$ non-Abelian orbifold of the \SO{32} heterotic string. We split it for
         the untwisted $U$ and twisted $T_{[\vartheta]}$, $T_{[\omega]}$ sectors.
         We have taken into account the multiplicities due to the 4 and 9 fixed points
         in the $T_{[\vartheta]}$ and $T_{[\omega]}$ sectors, respectively.
         \label{tab:S3spectrum}}
 \end{center}
\end{table}

%% file: results.tex
 \section{Other orbifold geometries}
\label{sec:results}

Let us now present the results we find by implementing our method for the orbifolds defined
by the point groups $D_{4}$ and $(\Z{4}\x\Z{2})\rtimes\Z{2}$.
As general features, in these cases, we find anomaly-free spectra, a number $\rep{26}$-plets compatible
with table~\ref{tab:n26summary} (which was obtained from topological arguments),
and rank reduction for the corresponding unbroken 4D gauge groups.

\subsection[D4 orbifold]{\boldmath$D_{4}$ orbifold\unboldmath}

We work with the representation of the $D_{4}$ group given by the generators
\begin{equation}
\vartheta=\begin{pmatrix}
1 & 0 & 0 & 0 & 0 & 0\\
0 & -1 & 0 & 0 & 0 & 0\\
0 & 0 & 1 & 0 & 0 & 0\\
0 & 0 & 0 & -1 & 0 & 0\\
0 & 0 & 0 & 0 & -1 & 0\\
0 & 0 & 0 & 0 & 0 & -1
\end{pmatrix},\qquad
\omega=\begin{pmatrix}
0 & -1 & 0 & 0 & 0 & 0\\
-1 & 0 & 0 & 0 & 0 & 0\\
0 & 0 & 0 & 1 & 0 & 0\\
0 & 0 & 1 & 0 & 0 & 0\\
0 & 0 & 0 & 0 & -1 & 0\\
0 & 0 & 0 & 0 & 0 & -1
\end{pmatrix}.
\end{equation}
The basis for its compatible lattice reads
\begin{equation}
\begin{split}
e'_{3}&=\left(1,0,0,0,0,0\right)^{\operatorname{T}}, \qquad e'_{4}=\left(0,1,0,0,0,0\right)^{\operatorname{T}},\\
e'_{5}&=\left(0,0,1,0,0,0\right)^{\operatorname{T}}, \qquad e'_{6}=\left(0,0,0,1,0,0\right)^{\operatorname{T}},\\
e'_{7}&=\left(0,0,0,0,1,0\right)^{\operatorname{T}}, \qquad e'_{8}=\left(0,0,0,0,0,1\right)^{\operatorname{T}}.
\end{split}
\end{equation}

Through an orthogonal transformation, $\vartheta$ is  
\begin{equation}
\vartheta=\begin{pmatrix}
-1 & 0 & 0 & 0 & 0 & 0\\
0 & 1 & 0 & 0 & 0 & 0\\
0 & 0 & 1 & 0 & 0 & 0\\
0 & 0 & 0 & -1 & 0 & 0\\
0 & 0 & 0 & 0 & -1 & 0\\
0 & 0 & 0 & 0 & 0 & -1
\end{pmatrix},
\end{equation}
while, through a different transformation, we have that $\omega$, $\vartheta\omega$ 
and $\vartheta\omega\vartheta\omega$ are   
\begin{equation}
\omega=\begin{pmatrix}
1 & 0 & 0 & 0 & 0 & 0\\
0 & -1 & 0 & 0 & 0 & 0\\
0 & 0 & 1 & 0 & 0 & 0\\
0 & 0 & 0 & -1 & 0 & 0\\
0 & 0 & 0 & 0 & -1 & 0\\
0 & 0 & 0 & 0 & 0 & -1
\end{pmatrix}, 
\end{equation}
and
\begin{equation}
\vartheta\omega=\begin{pmatrix}
0 & 1 & 0 & 0 & 0 & 0\\
-1 & 0 & 0 & 0 & 0 & 0\\
0 & 0 & 0 & -1 & 0 & 0\\
0 & 0 & 1 & 0 & 0 & 0\\
0 & 0 & 0 & 0 & 1 & 0\\
0 & 0 & 0 & 0 & 0 & 1
\end{pmatrix},\qquad 
\vartheta\omega\vartheta\omega = 
\begin{pmatrix}
-1& 0 & 0 & 0 & 0 & 0\\
0 & -1 & 0 & 0 & 0 & 0\\
0 & 0 & -1 & 0 & 0 & 0\\
0 & 0 & 0 & -1 & 0 & 0\\
0 & 0 & 0 & 0 & 1 & 0\\
0 & 0 & 0 & 0 & 0 & 1
\end{pmatrix}.
\end{equation}
Thus, the relevant Cartan basis are 
\begin{equation} \label{eq:CartanB_D4}
H_{\vartheta}=\left\{ J_{3,6},\,J_{4,5},\,J_{7,8} \right\},\quad H_{\omega}=\left\{ J_{3,5},\,J_{4,6},\,J_{7,8} \right\},\quad 
H_{\vartheta\omega}=\left\{ J_{3,4},\,J_{5,6},\,J_{7,8} \right\}. 
\end{equation}
Therefore, the representative elements of each conjugacy class are 
\begin{align}\label{eq:D4ccExp} 
\vartheta&=\operatorname{exp}\left[ \frac{2\pi\I}{2}(J_{3,6}-J_{7,8}) \right],&
\omega&=\operatorname{exp}\left[ \frac{2\pi\I}{2}(J_{4,6}-J_{7,8}) \right],\\
\vartheta\omega&=\operatorname{exp}\left[ \frac{2\pi\I}{4}(-J_{3,4}+J_{5,6}) \right],&
\vartheta\omega\vartheta\omega&=\operatorname{exp}\left[ \frac{2\pi\I}{2}(-J_{3,4}+J_{5,6}) \right].
\end{align}
Hence, the twists vectors in their corresponding basis are  
\begin{align}
v_{\vartheta}&=\left( 0,\frac{1}{2},0,-\frac{1}{2} \right),& 
v_{\omega}&=\left( 0,0,\frac{1}{2},-\frac{1}{2} \right),\\
v_{\vartheta\omega}&=\left( 0,-\frac{1}{4},\frac{1}{4},0 \right),&  
v_{\vartheta\omega\vartheta\omega}&=\left( 0,-\frac{1}{2},\frac{1}{2},0 \right).
\end{align}
Due to the standard embedding, the shift vectors are
\begin{align}
V_{\vartheta}&=\left( \frac{1}{2},0,-\frac{1}{2},0^{13} \right),& 
V_{\omega}&=\left( 0,\frac{1}{2},-\frac{1}{2},0^{13}  \right),\\ 
V_{\vartheta\omega}&=\left( -\frac{1}{4},\frac{1}{4},0,0^{13}  \right),&  
V_{\vartheta\omega\vartheta\omega}&=\left( -\frac{1}{2},\frac{1}{2},0,0^{13}  \right). 
\end{align}

Since the Cartan bases for the various sectors~\eqref{eq:CartanB_D4} share a common element, $J_{7,8}$, it is clear that such \SO{6}
generator commutes with every $D_{4}$ conjugacy class and it happens to be the only linear combination of \SO{6} generators 
with this property. Even more, it is the generator of a $\U{1}$ symmetry, so the 4D gauge group that we find in this case is 
\begin{equation}\label{eq:D4gg}
\SO{32} ~\to~ \U{1}\times\SO{26} ~=:~ G_{D_4}\,.
\end{equation} 
Since $\operatorname{rank}(G_{D_{4}})=14<16$, we conclude that we achieved rank reduction. 

Now, in order to complete our analysis, applying the techniques described in section~\ref{sec:spectrum}, we find the spectrum presented 
in table~\ref{tab:D4spectrum}. The relevant centralizers for this procedure are 
\begin{equation}
\mathcal{C}_{D_{4}}(g) \cong 
\begin{cases}
\Z{2}\times\Z{2} & g\in\{ \vartheta,\omega \}, \\ 
\Z{4} &  g= \vartheta\omega, \\ 
D_{4} & g=\vartheta\omega\vartheta\omega.
\end{cases}
\end{equation}
Notice that, even if some centralizers are isomorphic, say $\mathcal{C}_{D_{4}}(\vartheta)\cong\mathcal{C}_{D_{4}}(\omega)$, 
the elements that constitute them as subsets of $D_{4}$ are significantly different since 
\begin{equation}
\mathcal{C}_{D_{4}}(\vartheta) = \{ e,\,\vartheta,\,(\vartheta\omega)^{2},\,\vartheta(\vartheta\omega)^{2} \} \neq 
 \{ e,\,\omega,\,(\vartheta\omega)^{2},\,\omega(\vartheta\omega)^{2} \} = \mathcal{C}_{D_{4}}(\omega). 
\end{equation}
Thus, the respective orbifold projections have to be done carefully. Following our procedure, we find that the number of $\rep{26}$ representations
per sector agrees exactly with our findings in table~\ref{tab:n26summary}. 

\begin{table}[t]
\begin{center}
\begin{tabular}{|c|c|c|c|c|c|}
\hline 
 $\SO{26}\times\U{1}$ irrep & $U$ & $T_{\left[ \vartheta \right]} $ & $T_{\left[ \omega \right]} $ & $T_{\left[ \vartheta\omega \right]}$ & 
$T_{\left[ \vartheta\omega\vartheta\omega \right]}$   \\
\hline 
 $\rep{26}_{0}$ & 2 & 4 & 16 & 8 & 10  \\ 
\hline
$\rep{26}_{1}$ & 1 & 0 & 0 & 0 & 0  \\ 
\hline
$\rep{26}_{-1}$ & 1 & 0 & 0 & 0 & 0  \\ 
\hline
$\rep{1}_{0}$ & 2 & 16 & 0 & 0 & 20  \\ 
\hline
$\rep{1}_{-1}$ & 2 & 16 & 16 & 8 & 20  \\ 
\hline
$\rep{1}_{1}$ & 2 & 16 & 16 & 8 & 20  \\ 
\hline
\end{tabular}

\caption{Massless matter spectrum of the $D_{4}$ orbifold of the \SO{32} heterotic string allocated by sectors in the untwisted $U$, and the twisted
$T_{[\vartheta]}$, $T_{[\omega]}$, $T_{[\vartheta\omega]}$ sectors. Since there are 4, 16, 4 and 10 fixed points in each of our sectors,
we have considered these as multiplicities of the states.}
\label{tab:D4spectrum}
\end{center}
\end{table}

\subsection[(Z4 x Z2) rtimes Z2 orbifold]{\boldmath$(\Z4\x\Z2) \rtimes \Z2$ orbifold\unboldmath}

This group has three generators: one of order 4, $\rho$, and two of order two,
$\vartheta$ and $\omega$, given by 
\begin{equation}
\rho=
\begin{pmatrix}
1 & 0 & 0 & 0 & 0 & 0\\
0 & 0 & 0 & 0 & 1 & 0\\
0 & 0 & 0 & -1 & 0 & 0\\
0 & 0 & 1 & 0 & 0 & 1\\
0 & -1 & 0 & 0 & 0 & -1\\
0 & 0 & 0 & 0 & 0 & 1
\end{pmatrix},
\end{equation}
and
\begin{equation}
\vartheta=\begin{pmatrix}
-1 & 0 & 0 & 0 & 0 & 0\\
0 & 0 & -1 & 0 & 0 & 0\\
0 & -1 & 0 & 0 & 0 & 0\\
0 & 0 & 0 & 0 & 1 & 0\\
0 & 0 & 0 & 1 & 0 & 0\\
0 & 0 & 0 & 0 & 0 & -1
\end{pmatrix},\qquad
\omega=\begin{pmatrix}
-1 & 0 & 0 & 0 & 0 & 0\\
-1 & 0 & 0 & 0 & 1 & 1\\
0 & 0 & 0 & 1 & 0 & 0\\
0 & 0 & 1 & 0 & 0 & 0\\
0 & 1 & 0 & 0 & 0 & 1\\
0 & 0 & 0 & 0 & 0 & -1
\end{pmatrix}.
\end{equation}
The basis of the compatible lattice with this representation for the generators is  
\begin{equation}
\begin{split}
e'_{3}=\left(1,0,0,0,0,0\right)^{\operatorname{T}}, \qquad e'_{4}&=\left(0,\sqrt{2},0,0,0,0\right)^{\operatorname{T}},\\
e'_{5}=\left(0,0,\sqrt{2},0,0,0\right)^{\operatorname{T}}, \qquad e'_{6}&=\left(0,0,0,\sqrt{2},0,0\right)^{\operatorname{T}},\\
e'_{7}=\left(0,0,0,0,\sqrt{2},0\right)^{\operatorname{T}},  \qquad 
e'_{8}&=\left(0,\frac{1}{\sqrt{2}},\frac{1}{\sqrt{2}},-\frac{1}{\sqrt{2}},\frac{1}{\sqrt{2}},1\right)^{\operatorname{T}}.
\end{split}
\end{equation}
This point group has eight different non-trivial conjugacy classes. Their representative elements are the
elements
\begin{equation}
\left\{ \rho,\,\vartheta,\,\omega,\,\vartheta\omega,\,\vartheta\rho,\,\omega\rho,\,\vartheta\omega\rho^{3},\,\rho^{2} \right\}.   
\end{equation}

Since this geometry is more complex, 5 different orthogonal transformations are needed to find the desired form for each representative 
element. 

We find that no linear combination of the \SO{6} generators commutes simultaneously with every conjugacy class. Therefore, the 4D
gauge group that we find in this case is 
\begin{equation}\label{eq:Z4Z2Z2gg}
   \SO{32}\to \SO{26} =: G_{(\Z{4}\x\Z{2})\rtimes\Z{2}}.
\end{equation} 
Note that $\operatorname{rank}\left( G_{(\Z{4}\x\Z{2})\rtimes\Z{2}} \right)=13<16$. This is the maximal rank reduction achievable
through standard embedding.

Naturally, we compute the spectrum presented in table~\ref{tab:Z4Z2Z2spectrum} with our previous techniques taking into account the 
centralizers for each conjugacy class. Such centralizers are 
\begin{equation}
\mathcal{C}_{(\Z{4}\times\Z{2})\rtimes\Z{2}}(g) = 
\begin{cases}
(\Z{4}\times\Z{2})\rtimes\Z{2} & g\in\{ e,\, \rho^{2} \}, \\ 
\Z{4}\times\Z{2} &  g\in\{ \rho,\,  \vartheta,\,  \vartheta\omega \}, \\ 
\Z{2}\times\Z{2} & g=\omega,\\ 
\Z{4} & g\in\{ \vartheta\rho,\, \omega\rho,\,\vartheta\omega\rho^{3} \}.
\end{cases}
\end{equation}
Again, for our calculations we must consider the difference of the centralizers viewed as sets. 

With this in mind, we obtain that the number of $\rep{26}$ in table~\ref{tab:Z4Z2Z2spectrum} is in agreement with our prior results in 
table~\ref{tab:n26summary}. Thus, our method is consistent for every geometry we have discussed. 
\begin{table}[t]
\begin{center}
\begin{tabular}{|c|c|c|c|c|c|c|c|c|c|}
\hline 
 $\SO{26}$ irrep & $U$ & $T_{\left[ \vartheta \right]} $ & $T_{\left[ \omega \right]}$ & $T_{\left[ \rho \right]}$ & $T_{\left[ \vartheta\omega \right]}$ & 
$T_{\left[ \vartheta\rho \right]}$ & $T_{\left[ \omega\rho \right]}$ & $T_{\left[ \vartheta\omega\rho^{3} \right]}$ &  $T_{\left[ \rho^{2} \right]}$  \\ 
\hline 
 $\rep{26}$ & 3 & 4 & 4 & 10 & 2 & 2 & 2 & 6 & 5 \\ 
\hline
 $\rep{1}$ & 3 & 8 & 8 & 40 & 8 & 8 & 8 & 24 & 20  \\ 
\hline
\end{tabular}

\caption{Massless matter spectrum of the non-Abelian $(\Z{4}\x\Z{2})\rtimes\Z{2}$ orbifold of the \SO{32} heterotic string. We distribute it by sectors:
untwisted $U$ and twisted $T_{[\rho]}$, $T_{[\vartheta]}$, $T_{[\omega]}$, $T_{[\vartheta\omega]}$, $T_{[\vartheta\rho]}$, $T_{[\omega\rho]}$, 
$T_{[\vartheta\omega\rho^{3}]}$, $T_{[\rho^{2}]}$. As before, multiplicities have been accounted for due to the 4, 4, 10, 2, 2, 2, 6, and 5 fixed points 
in each respective sector.}
\label{tab:Z4Z2Z2spectrum}
\end{center}
\end{table}

%% file: appendix.tex
\section{\boldmath \SO{32} heterotic weight lattices \unboldmath}
\label{app:Lattices}

The {\bf weight lattice of $\rep{\SO{8}}$} contains the 4D weight vectors of all representations of the group.
It is possible to build all \SO8 representations by tensor products of two of the three inequivalent
8-dimensional representations, which can be chosen as the vector $\rep8_v$ and the spinor $\rep8_s$
representations (the conjugate spinor $\rep8_c$ is included in $\rep8_v\otimes\rep8_s$).
Hence, choosing a basis for the weights of these representations suffices to build the whole weight lattice.
Our choice reads
\begin{subequations}
\label{eq:sols_qsquare1}
\begin{align}
   \rep8_v &\sim (\pm1,0,0,0) \text{ and its permutations}\,,\\ 
   \rep8_s &\sim (\pm\nicefrac12,\pm\nicefrac12,\pm\nicefrac12,\pm\nicefrac12) \text{ with even \# of plus signs}\,.
\end{align}
\end{subequations}
This implies that the whole \SO8 weight lattice can be given by the
\begin{subequations}
\begin{align}
   \text{vector conjugacy class} & :~ (n_1,n_2,n_3,n_4) &\text{ with } \sum n_i = \text{odd}\,,\\
   \text{spinor conjugacy class} & :~ (n_1+\nicefrac12,n_2+\nicefrac12,n_3+\nicefrac12,n_4+\nicefrac12) &\text{ with } \sum n_i = \text{even}\,.
\end{align}
\end{subequations}
where $n_i\in\Z{}$. Note that if $q$ denotes elements of the \SO8 weight lattice such that $q^2=1$, then
the solutions build the fundamental vector and spinor representations given in eqs.~\eqref{eq:sols_qsquare1}.

On the other hand, the {\bf weight lattice of $\rep{\mathrm{Spin}(32)/\Z2}$} can be chosen to be spanned
by the weights associated with the scalar and the spinor conjugacy classes, given by
\begin{subequations}
\begin{align}
   \text{scalar conjugacy class} & :~ (n_1,n_2,\ldots,n_{16}) \,,\\
   \text{spinor conjugacy class} & :~ (n_1+\nicefrac12,n_2+\nicefrac12,\ldots,n_{16}+\nicefrac12)\,.
\end{align}
\end{subequations}
where $n_i\in\Z{}$ and $\sum n_i = \text{even}$. Notice that if $p$ denotes elements of the weight
lattice of $\mathrm{Spin}(32)/\Z2$ satisfying $p^2=2$, then they take the form of all 480 permutations
of the element $(\pm1,\pm1,0,\ldots,0)$, arising from the scalar conjugacy class.
These can be interpreted as the non-trivial weights of the
adjoint representation $\rep{496}$ (i.e.\ the weights of the simple roots) of \SO{32}, 
which is the representation under which the gauge bosons
of a Yang-Mills theory based on \SO{32} transform.

\section{\boldmath Invariance of $S_3$ singlets under a change of basis \unboldmath}
\label{app:GT}

In order to provide solid arguments that support the formalism that we have developed, it is necessary to show that given
a discrete group $P$ its irreducible representations, specially its singlets, are mapped correctly under the change 
of basis transformations that we proposed in section~\ref{sec:spectrum}. 
With this requirement in mind, we show that this fact holds for $P=S_{3}$ as a working example. 

We begin by consider the presentation and the form of the generators of $S_{3}$ used in section~\ref{sec:spectrum}. 
Therefore, the generators $\vartheta$ and $\omega$ have the form 
\begin{equation}\label{eq:S3gens}
\vartheta=\begin{pmatrix}
1 & -1 & 0 & 0 & 0 & 0\\
0 & -1 & 0 & 0 & 0 & 0\\
0 & 0 & 1 & -1 & 0 & 0\\
0 & 0 & 0 & -1 & 0 & 0\\
0 & 0 & 0 & 0 & -1 & 0\\
0 & 0 & 0 & 0 & 0 & -1
\end{pmatrix},\qquad
\omega=\begin{pmatrix}
0 & -1 & 0 & 0 & 0 & 0\\
1 & -1 & 0 & 0 & 0 & 0\\
0 & 0 & -1 & 1 & 0 & 0\\
0 & 0 & -1 & 0 & 0 & 0\\
0 & 0 & 0 & 0 & 1 & 0\\
0 & 0 & 0 & 0 & 0 & 1
\end{pmatrix},
\end{equation}

For our purposes it is enough to consider the $2D$ case. So, we extract two blocks that according to GAP~\cite{GAP4} 
are indeed generators of $S_{3}$, say 
\begin{equation}\label{eq:S3_2Dgens}
\vartheta_{2D}=\begin{pmatrix}
1 & -1 \\
0 & -1 \\
\end{pmatrix},\qquad
\omega_{2D}=\begin{pmatrix}
0 & -1 \\
1 & -1 \\
\end{pmatrix}.
\end{equation}
We proceed considering two doublets, and their tensor product 
\begin{equation}\label{eq:doublets}
x=\begin{pmatrix}
x_{1} \\
x_{2} \\
\end{pmatrix},\qquad
y=\begin{pmatrix}
y_{1} \\
y_{2} \\
\end{pmatrix},\qquad
x\otimes y = \begin{pmatrix}
x_{1} y_{1} & x_{1} y_{2}  \\
x_{2} y_{2} & x_{2} y_{2} \\
\end{pmatrix}.
\end{equation} 
Now, we let the generators~\eqref{eq:S3_2Dgens} act on $x\otimes y$ to uncover its transformation under $\vartheta_{2D}$
\begin{equation}
\begin{split}
x_{1}y_{1}&\xrightarrow{\vartheta}  x_{1}y_{1} - x_{2}y_{1},\\
x_{1}y_{2}&\xrightarrow{\vartheta}  - x_{1}y_{1} - x_{1}y_{2} + x_{2}y_{1} + x_{2}y_{2},\\
x_{2}y_{1}&\xrightarrow{\vartheta}  - x_{2}y_{1},\\
x_{2}y_{2}&\xrightarrow{\vartheta} x_{2}y_{1} + x_{2}y_{2},
\end{split}
\end{equation}
and under $\omega_{2D}$
\begin{equation}
\begin{split}
x_{1}y_{1}&\xrightarrow{\omega}  x_{2}y_{2} + x_{2}y_{1},\\
x_{1}y_{2}&\xrightarrow{\omega}  - x_{2}y_{1},\\
x_{2}y_{1}&\xrightarrow{\omega}   - x_{1}y_{1} - x_{1}y_{2} + x_{2}y_{1} + x_{2}y_{2},\\
x_{2}y_{2}&\xrightarrow{\omega} x_{1}y_{1} - x_{2}y_{1}.
\end{split}
\end{equation}
Therefore, we observe that the singlet $\rep{1}$ under both $\vartheta_{2D}$ and $\omega_{2D}$ is  
\begin{equation}
\rep{1} = x_{1}y_{1} +  x_{2}y_{2}.
\end{equation}
Let us now consider the relevant part of the change of the basis transformation~\eqref{eq:S3Q},
\begin{equation}
Q_{2D}=\begin{pmatrix}
1 & -\frac{1}{2} \\
0 & \frac{\sqrt{3}}{2}
\end{pmatrix}.
\end{equation} 
Under $Q_{2D}$, $\vartheta_{2D}$ and $\omega_{2D}$ take the form
\begin{equation}\label{eq:S3_2Dgens_diag}
\vartheta'_{2D}=\begin{pmatrix}
1 & 0 \\
0 & -1 \\
\end{pmatrix},\qquad
\omega'_{2D}=\begin{pmatrix}
-\frac{1}{2} & -\frac{\sqrt{3}}{2} \\
\frac{\sqrt{3}}{2} & -\frac{1}{2} \\
\end{pmatrix},
\end{equation}
while the tensor product of the doublets transforms as
\begin{equation}
x\otimes y\, \xrightarrow{Q} \,(x\otimes y)':=
\begin{pmatrix}
(x_1y_1)' & (x_1y_2)'\\
(x_2y_1)' & (x_2y_2)'
\end{pmatrix}
=
\begin{pmatrix}
x_{1} y_{1} -\frac{x_{2}y_{1}}{2} & \frac{1}{2\sqrt{3}} (2x_{1}-x_{2})(y_{1}+2y_{2}) \\
\frac{\sqrt{3}}{2} x_{2} y_{1} & \frac{x_{2}y_{1}}{2} + x_{2}y_{2} \\
\end{pmatrix}. 
\end{equation}
Applying $\vartheta'_{2D}$ and $\omega'_{2D}$ on the $(x\otimes y)'$, we find that
its components transform as
\begin{equation}
\begin{split}
(x_{1}y_{1})'&\xrightarrow{\vartheta'}  (x_{1}y_{1})' = x_{1} y_{1} -\frac{x_{2}y_{1}}{2},\\
(x_{1}y_{2})'&\xrightarrow{\vartheta'}  -(x_{1}y_{2})' = -\frac{\sqrt{3}}{6} (2x_{1}-x_{2})(y_{1}+2y_{2}),\\
(x_{2}y_{1})'&\xrightarrow{\vartheta'}  -(x_{2}y_{1})' = -\frac{\sqrt{3}}{2} x_{2} y_{1},\\
(x_{2}y_{2})'&\xrightarrow{\vartheta'} (x_{2}y_{2})' = \frac{x_{2}y_{1}}{2} + x_{2}y_{2},
\end{split}
\end{equation}
and
\begin{equation}
\begin{split}
(x_{1}y_{1})'&\xrightarrow{\omega'} \frac{1}{2}(x_{1}y_{1} + x_{1}y_{2} + x_{2}y_{1}+x_{2}y_{2}),\\
(x_{1}y_{2})'&\xrightarrow{\omega'} -\frac{\sqrt{3}}{6}(x_{1}y_{1} - x_{1}y_{2} + x_{2}y_{1} - x_{2}y_{2}),\\
(x_{2}y_{1})'&\xrightarrow{\omega'} -\frac{\sqrt{3}}{2}(x_{1}y_{1} + x_{1}y_{2} - x_{2}y_{1} - x_{2}y_{2}),\\
(x_{2}y_{2})'&\xrightarrow{\omega'} \frac{1}{2}(x_{1}y_{1} - x_{1}y_{2} - x_{2}y_{1} + x_{2}y_{2}).
\end{split}
\end{equation}
Therefore, it is evident that the singlet in the new basis is 
\begin{equation}
\rep{1} = (x_{1}y_{1})' +  (x_{2}y_{2})' = x_{1}y_{1} + x_{2}y_{2},
\end{equation}
i.e.\ the singlet $\rep{1}$ is invariant under our transformation $Q$.

\section{Block diagonalization of the orbifold point group}
\label{app:BDdetails}

As described in section~\ref{sec:spectrum}, our task is to find a basis for every conjugacy class of
the point groups in table~\ref{tab:naoN1}, such that its generator has diagonal form. That is, if
$[\theta]$ denotes the conjugacy class of a point group element with twist vector
$v_{\theta}=(0,v^{1},v^{2},v^{3})$, then we seek that $\theta$ takes the form
\begin{equation} \label{eq:bdform}
\theta \rightarrow 
\begin{pmatrix}
\operatorname{Rot}(2\pi v^{1}) & 0_{2} & 0_{2}\\
0_{2} &\operatorname{Rot}(2\pi v^{2}) & 0_{2}\\
0_{2} & 0_{2} &\operatorname{Rot}(2\pi v^{3})
\end{pmatrix}. 
\end{equation}
Hence, this problem can be understood in terms of block diagonalization of $\theta$, with a specific form 
for every block. 

The method that we have developed to achieve block diagonalization 
is entirely based on the main theorem presented in~\cite{Eisenfeld:1976}. Roughly, this theorem provides us 
criteria to determine if a given square $n$-dimensional matrix $A$ is similar to a block diagonal matrix,
say $D$, given by 
\begin{equation}\label{eq:Dmatrix} 
D= 
\begin{pmatrix}
A_{1} & 0_{qp}\\
0_{pq} & A_{2}
\end{pmatrix}, 
\end{equation} 
where $A_{1}$ is a square $p$-dimensional matrix and $A_{2}$ is a $q$-dimensional one, with $n=p+q$.

The procedure starts with the decomposition of the matrix $A$ in four submatrices $E$, $F$, $H$ and $G$ 
according to
\begin{equation}\label{eq:Amatrix}
A=
\begin{pmatrix}
E & F\\
G & H 
\end{pmatrix},
\end{equation}
where $E$, $H$, $F$ and $G$ are $p\times p$, $q\times q$, $q\times p$ and $p\times q$ matrices,
respectively. 

Now, as stated in~\cite{Eisenfeld:1976}, the matrix \eqref{eq:Amatrix} is similar to \eqref{eq:Dmatrix}
if there exist two matrices $R$ and $X$ that satisfy
\begin{equation} \label{eq:RX}
R (E+FR)~=~G+HR\,,\quad (E+FR)X-X(H-RF)~=~-F\,.
\end{equation}
In such a case, as proved in~\cite{Eisenfeld:1976}, the transformation matrix that relates
the matrices \eqref{eq:Amatrix} and \eqref{eq:Dmatrix} is given by
\begin{equation} \label{eq:Wmat}
W= 
\begin{pmatrix}
\Id_{p} & X \\
R & RX + \Id_{q}
\end{pmatrix},
\end{equation} 
and the explicit form of the $D$ matrix is 
\begin{equation}
D=
W^{-1} A W = 
\begin{pmatrix}
E+FR & 0_{qp} \\
0_{pq} & H-RF
\end{pmatrix}.
\end{equation} 
With this result in hand, given an element $\theta$ with twist vector 
$v_{\theta}=(0,v^{1},v^{2},v^{3})$, we must find a transformation $T$,
such that its similar matrix $D$ has an even more restricted form as discussed in section~\ref{sec:spectrum},
we would like to  find a transformation $Q$ that gives us the specific form 

\begin{equation} \label{eq:bdformrot} 
\theta\rightarrow
\begin{pmatrix}
\operatorname{Rot}(2\pi v^{1}) & 0_{2} & 0_{2}\\
0_{2} &\operatorname{Rot}(2\pi v^{2}) & 0_{2}\\
0_{2} & 0_{2} &\operatorname{Rot}(2\pi v^{3})
\end{pmatrix}. 
\end{equation} 

Now, as reported in~\cite{Fischer:2012qj}, we have 35 different point groups that give 
rise to 331 inequivalent orbifold geometries. From these, we observe that 263 of
them are associated with 18 point groups. Every conjugacy class of these 18 groups share the
property of having two invariant directions, i.e.\ every conjugacy class representative has
a twist vector $v=\left(0, v^{1},v^{2},v^{3} \right)$, such that $v^{i}=0$ for some 
$i\in\{1,2,3\}$. Due to this fact, it is relevant to focus our efforts on this simpler
case, i.e.\ our original problem gets thereby reduced to finding a $W$, such that
\begin{equation} 
\theta \xrightarrow{W^{-1} \theta W} 
\begin{pmatrix}
\mathbb{I}_{2} & 0_{2} & 0_{2}\\
0_{2} &\operatorname{Rot}(2\pi v^{2}) & 0_{2}\\
0_{2} & 0_{2} &\operatorname{Rot}(2\pi v^{3})
\end{pmatrix}.
\end{equation} 

As a consequence, the first thing to do is to find the geometries that have two eigenvectors with eigenvalue $1$.
We achieve a reduction of the problem: using a Gramm-Schmidt algorithm to move to the basis that
has those eigenvectors as its first two elements, we find that our starting element $\theta$
can be written as
\begin{equation} \label{eq:gstheta} 
\theta=
\begin{pmatrix}
\Id_{2\times2} & 0_{2\times 4}\\
0_{4\times 2} &B_{4\times4}
\end{pmatrix}.
\end{equation} 
Therefore, we have to solve the eqs.~\eqref{eq:RX} for the matrix $B$ in eq.~\eqref{eq:gstheta}, and then
to calculate the entries of the matrix $Q$ that lead to the expression in eq.~\eqref{eq:bdformrot}.

This method allows us to work with the following geometries
\begin{equation}\label{eq:solvablegroups}
\begin{split}
\tilde{P} = &\{
S_{3},\,D_{4},\,A_{4},\, 
D_{6},\,(\mathbb{Z}_{4}\times\mathbb{Z}_{2})\rtimes\mathbb{Z}_{2},\,\mathbb{Z}_{4}\rtimes S_{3},\,S_{4},\\ &
(\mathbb{Z}_{4}\times\mathbb{Z}_{4})\rtimes\mathbb{Z}_{2},\, \mathbb{Z}_{3}\times(\mathbb{Z}_{3}\rtimes
\mathbb{Z}_{4}),\, \Delta(27),\,\Delta(54),\,\Delta(96)
\}. 
\end{split}
\end{equation}
This list represents one third of the inequivalent point groups from table~\ref{tab:naoN1}. However, these 
point groups give rise to 219 non-Abelian orbifold geometries out of all 331 admissible geometries.

Now, even if eq.~\eqref{eq:solvablegroups} represents the majority of the geometries, we have found that the matrix
$W$ is orthogonal only for a subset of all admissible geometries:
\begin{equation} \label{eq:finallist}
\tilde{P}' ~=~ \left\{
S_{3},\,D_{4},\,(\mathbb{Z}_{4}\times\mathbb{Z}_{2})\rtimes\mathbb{Z}_{2}
\right\}. 
\end{equation} 
Note that the orthogonality condition is, in general, not mandatory to compute the matter spectrum.
However, we impose it here because our method to arrive at the massless matter spectrum of non-Abelian orbifolds,
as presented in section~\ref{sec:spectrum}, crucially relies on it.

\pagebreak
\begin{landscape}
\section{\boldmath Non-trivial representations of \SO{26} in non-Abelian orbifolds \unboldmath}
\label{app:26s}

In this section, using the results of ref.~\cite{Fischer:2013qza} for the \E8\x\E8
heterotic string, we determine the number of fundamental representations $\rep{26}$ of $\SO{26}$ in each
of the non-Abelian orbifolds of the \SO{32} heterotic string with standard gauge embedding.

\scriptsize
\begin{longtable}{|c|c|c|c|c|}
\hline
\multirow{2}{*}{$P$} & \multirow{2}{*}{$\mathbb{Z}-$class} &  \multirow{2}{*}{affine class} & Generators of $P$ & \\
\cline{4-5}
& & & number of $\rep{26}$ from $U$ and $T$ sectors & total $\rep{26}$ \\
\hline
\endhead
\hline
\multicolumn{5}{r@{}}{continued \ldots}\\
\endfoot
\endlastfoot
\multirow{6}{*}{$S_{3}$} & \multirow{2}{*}{1} & \multirow{2}{*}{1} & $(\vartheta,0),\,(\omega,0)$ & \\
& &  & $(4)U\, + (8)T_{[\vartheta]}+ (18)T_{[\omega]}$ & 30 \\
\cline{3-5}
 &  & \multirow{2}{*}{2} & $(\vartheta,0),\,(\omega,\frac{1}{3}e_{5})$ & \\
 & &  & $(4)U\, + (8)T_{[\vartheta]}$ & 12 \\
\cline{2-5}
 & \multirow{2}{*}{2} & \multirow{2}{*}{1} & $(\vartheta,0),\,(\omega,0)$ & \\
& &  & $(4)U\, + (8)T_{[\vartheta]}+ (18)T_{[\omega]}$ & 30 \\
\cline{3-5}
 &  & \multirow{2}{*}{2} & $(\vartheta,0),\,(\omega,\frac{1}{3}e_{5})$ & \\
 & &  & $(4)U\, + (8)T_{[\vartheta]}$ & 12 \\
\end{longtable}
\end{landscape}
\normalsize